\documentclass[journal,10pt,letter]{IEEEtran}

\usepackage[utf8]{inputenc}
\usepackage{cite}
\usepackage{amsmath,amssymb,amsfonts}
\usepackage{algorithmic}
\usepackage{graphicx}
\usepackage{textcomp}
\usepackage[caption=false]{subfig}
\usepackage[justification=centering]{caption}
\usepackage[ruled,vlined]{algorithm2e}
\usepackage{epsfig}
\usepackage{float}
\usepackage{color}
\usepackage{balance}
\usepackage{bbm}
\usepackage{textcomp}
\usepackage{enumitem}

\DeclareMathOperator{\diag}{\mathrm{diag} \, }

\newsavebox\myboxA
\newsavebox\myboxB
\newlength\mylenA

\def\mD{\mbox{$\mathbf{D}$}}

\def\mH{\mbox{$\mathbf{H}$}}

\def\mSigma{\mbox{$\mathbf{\Sigma} \kern .08em$}}
\def\mLambda{\mbox{$\mathbf{\Lambda} \kern .08em$}}

\def\b0{\text{\mbox{\boldmath $0$}}}

\def\bg{\text{\mbox{\boldmath $g$}}}
\def\bh{\text{\mbox{\boldmath $h$}}}
\def\bh{\text{\mbox{\boldmath $h$}}}

\def\bz{\text{\mbox{\boldmath $z$}}}

\def\bv{\text{\mbox{\boldmath $v$}}}

\newcommand{\qedsymbol}{\hspace{\fill}\rule{1.5ex}{1.5ex}}

\newenvironment{rcases}
  {\left.\begin{aligned}}
  {\end{aligned}\right\rbrace}

\floatstyle{ruled}
\newfloat{algorithm}{tbp}{loa}
\providecommand{\algorithmname}{Algorithm}
\floatname{algorithm}{\protect\algorithmname}

\IEEEoverridecommandlockouts

\begin{document}
\title{Dynamic Edge Computing empowered by Reconfigurable Intelligent Surfaces \vspace{.1cm}}

\author{Paolo Di Lorenzo~\IEEEmembership{ Senior Member,~IEEE}, Mattia Merluzzi,~\IEEEmembership{ Member,~IEEE},\\ 
Emilio Calvanese Strinati,~\IEEEmembership{ Member,~IEEE}, Sergio Barbarossa,~\IEEEmembership{ Fellow,~IEEE} \thanks{ P.~Di Lorenzo and S.~Barbarossa are with: 1) the Department of Information Engineering, Electronics, and Telecommunications of Sapienza University, via Eudossiana 18, 00184 Roma, Italy; 2) Consorzio Nazionale Interuniversitario per le Telecomunicazioni (CNIT), Parma, Italy.
E-mail: paolo.dilorenzo@uniroma1.it, sergio.barbarossa@uniroma1.it. M.~Merluzzi and E. Calvanese Strinati are with Univ. Grenoble Alpes, CEA, Leti, F-38000 Grenoble, France, France. \\Email: mattia.merluzzi@cea.fr, emilio.calvanese-strinati@cea.fr. The work of Di Lorenzo, Calvanese Strinati, and Barbarossa was supported by the European Union H2020 RISE-6G project no. 101017011. Barbarossa's work was also supported by MIUR under the PRIN Liquid-Edge contract.} \vspace{-.7cm}}

\maketitle

\begin{abstract}
In this paper, we propose a novel algorithm for energy-efficient, low-latency dynamic mobile edge computing (MEC), in the context of beyond 5G networks endowed with Reconfigurable Intelligent Surfaces (RISs). In our setting, new computing requests are continuously generated by a set of devices and are handled through a dynamic queueing system. Building on stochastic optimization tools, we devise a dynamic learning algorithm that jointly optimizes the allocation of radio resources (i.e., power, transmission rates, sleep mode and duty cycle), computation resources (i.e., CPU cycles), and RIS reflectivity parameters (i.e., phase shifts), while guaranteeing a target performance in terms of average end-to-end (E2E) delay. The proposed strategy is dynamic, since it performs a low-complexity optimization on a per-slot basis while dealing with time-varying radio channels and task arrivals, whose statistics are unknown. The presence and optimization of RISs helps boosting the performance of dynamic MEC, thanks to the capability to shape and adapt the wireless propagation environment. Numerical results assess the performance in terms of service delay, learning, and adaptation capabilities of the proposed strategy for RIS-empowered MEC.
\end{abstract}


\vspace{-.4cm}

\section{Introduction}

With the advent of beyond 5G networks \cite{ahmadi20195g,6Gstrinati}, mobile communication systems are evolving from a pure communication framework to enablers of a plethora of new services (including \textit{verticals}), such as Industry 4.0, Internet of Things (IoT), and autonomous driving, building on the tight integration of communication, computation, caching, and control \cite{Barbarossabook2018,Ndikumama19,Merl2021EML}. These new services have very different requirements and they generally involve massive data processing within low end-to-end delays. Among several technology enablers at different layers (e.g., artificial intelligence, network function virtualization, millimeter-wave communications), a prominent role will be played by Mobile Edge Computing (MEC), whose aim is to move cloud functionalities (e.g., computing and storage resources) at the edge of the wireless network to avoid the relatively long and highly variable delays necessary to reach centralized clouds. MEC-enabled networks allow User Equipments (UEs), to offload computational tasks to nearby processing units or Edge Servers (ESs), typically placed close to Access Points (APs), in order to run the computation on the UEs' behalf. However, since ESs have much smaller computation capabilities than the servers in the cloud, the available resources (i.e., radio, computation, energy) have to be properly managed to provide the end users with a satisfactory Quality  of  Service  (QoS).  In  particular, since the end-to-end delay includes a communication time and a computation time, the resources available at the wireless network edge must be managed jointly, learning over time the best joint resource allocation in a dynamic and data-driven fashion.

\textit{Related works on MEC:} There is a wide literature on computation offloading, aimed at jointly optimizing communication and computation resources in both static and dynamic MEC scenarios  \cite{barbarossa2014communicating,you2017,Mao2016,Mao2017,Chen2019,Merluzzi2020URLLC,HanChen2020,merluzzi2020d}. Recent surveys on the topic appear also in \cite{mach2017mobile} and \cite{mao2017survey}. A possible classification of computation offloading problems is between {\it static} and {\it dynamic} strategies. The static formulation deals with short time applications, in which mobile users send a single computation request, typically specifying also a service time \cite{barbarossa2014communicating,munoz2014optimization,you2017,zhao2017energy}. Conversely, in a dynamic scenario, the application continuously generates data to be processed, sometimes with an unknown rate. A typical example could be the transmission of a video, recorded by a mobile device, to be processed at the ES side for pattern recognition or anomaly detection. The dynamic formulation is also useful to handle users' mobility, which is a central problem in mobile networks and becomes even more central in a MEC environment, where mobility may require handover mechanisms involving both radio access points (APs) and ESs. In \cite{huang2012dynamic}, a dynamic formulation is proposed, with a strategy based on Lyapunov optimization in a cloud computing framework. In \cite{Merluzzi2020URLLC}, a Lyapunov based strategy is proposed, for the joint optimization of radio and computation resources, to minimize the users' energy consumption under E2E delay constraints. In \cite{Chen2019}, the authors investigate a scenario with multiple APs and edge servers, where an assignment strategy based on matching theory is proposed, coupled with the tools of Lyapunov optimization and Extreme Value Theory to control reliability. In \cite{merluzzi2020d}, a discontinuous mobile edge computing framework is proposed to minimize the energy consumption under latency constraints, considering a holistic approach that comprises UEs, APs, and ESs. In \cite{bi2021lyapunov}, a deep reinforcement learning strategy driven by Lyapunov optimization is proposed to enable stable offloading in dynamic MEC. Finally, in \cite{merluzzi2021wireless}, a dynamic resource allocation framework is proposed for edge learning, encompassing communication, computation, and inference/training aspects of the learning task.

\textit{RIS-empowered wireless networks:} All the aforementioned works considered the presence of a suitable wireless propagation environment to enable edge computing. However, moving toward millimeter wave (mmWave) communications (and beyond), poor channel conditions due to mobility, dinamicity of the environment, and blocking events,  might severely hinder the performance of MEC systems. In this context, a strong performance boost can be achieved with the advent of reconfigurable intelligent surfaces \cite{di2020smart,RISE-EUCNC}, which are artificial surfaces made of hundreds of nearly passive (sometimes, also active) reflective elements that can be programmed and controlled to realize dynamic transformations of the wireless propagation environment, both in indoor and outdoor scenarios. More precisely, an RIS is an array of backscatterers, where each element applies an individual phase-shift (and/or an amplitude and/or a polarization rotation) with which it backscatters an incident
wave \cite{di2019smart,mursia2020risma,zhang2020reconfigurable}, with the aim of creating a controllable reflected beam. RISs enable full programmability of the wireless propagation environment and dynamically create service \textit{boosted areas} where capacity, energy efficiency, and reliability can be dynamically traded to meet momentary and location dependent requirements \cite{Strinati2021reconfigurable}. Thus, RISs offer new opportunities to boost uplink and downlink capacities, and to counteract channel blocking effects in case of directive mmWave communications.

In the literature, several works have already investigated the optimization of RIS-empowered wireless communications. In \cite{huang2019reconfigurable},
a joint transmit power allocation and phase shift design was developed for an RIS-based multiuser system to maximize the energy efficiency. In \cite{wu2019intelligent}, the authors considered a downlink RIS-assisted multiuser communication system and studied a joint transmission and reflection beamforming problem to minimize the total transmit power. Also, very recently, a few papers exploited RISs to enhance the performance of MEC systems, jointly optimizing computing, communications and RIS parameters \cite{bai2020latency,chu2020intelligent,huang2021reconfigurable,hu2021reconfigurable}. In particular, in \cite{bai2020latency}, the authors propose a latency-minimization problem for multi-device scenarios, which optimizes the computation offloading volume, the edge
computing resource allocation, the multi-user detection matrix, and the RIS phase shifts, subject to a total edge computing capability. Reference \cite{chu2020intelligent} maximizes instead the number of processed bits for computation offloading, optimally designing the ES CPU frequency, the offloading time allocation, the transmit power of each device, and phase shifts of the RIS. Then, the work in \cite{huang2021reconfigurable} exploits RISs to maximize the performance of a machine learning task run at the edge server, acting jointly on radio parameters such as power of UEs, beamforming vectors of AP, and RIS parameters. Finally, reference \cite{hu2021reconfigurable} proposes optimization-based and data-driven solutions for RIS-empowered multi-user mobile edge computing, maximizing the total completed task-input bits of all UEs with limited energy budgets. All these previous works focus on a \textit{static} edge computing scenario. Conversely, in this paper we focus on a \textit{dynamic}  scenario, where UEs continuously generate data to be offloaded (e.g. a video stream for object detection), experiencing time varying context parameters (e.g. wireless channels conditions, data generation, server utilization, etc.).

\textit{Contributions of the paper:} In this work, we propose a novel algorithmic framework for energy-efficient, RIS-empowered dynamic mobile edge computing. To the best of our knowledge, this is the first contribution available in the literature in the context of dynamic MEC empowered by RISs, and extends the preliminary conference precursor in \cite{DiLo-RISMEC}. In our dynamic system model, at each time slot, new offloading requests are generated by the UEs, and are handled through a dynamic queueing system that accounts for both communication (i.e., uplink and downlink) and processing delays. In view of this dynamic MEC scenario, we devise a dynamic algorithm that learns over time the optimal radio parameters (i.e., powers, rates) for both UEs and AP (including also AP sleep mode and duty cycle), computation resources (i.e., CPU cycles) of the ES, and RIS reflectivity parameters (i.e., phase shifts), with the aim of enabling energy-efficient mobile edge computing with low end-to-end latency guarantees. The method hinges on Lyapunov stochastic optimization, and allocates resources in a dynamic fashion requiring only low-complexity operations at each slot (with semi-closed form expressions). Furthermore, the method does not require any prior knowledge of channel and data arrival statistics, and is able to learn and adapt in real-time to changes in the environment due to, e.g., mobility of UE's or channel blocking. In this way, Lyapunov stochastic optimization acts as a method that dynamically learns optimal control policies (i.e., resource allocation) over time in an online and data-driven fashion (i.e., observing channel and data realizations). Finally, we assess the performance of the proposed strategy through numerical simulations, illustrating how RISs help boosting the performance of MEC systems.

\begin{figure*}[t]
    \centering
    \includegraphics[width=11cm]{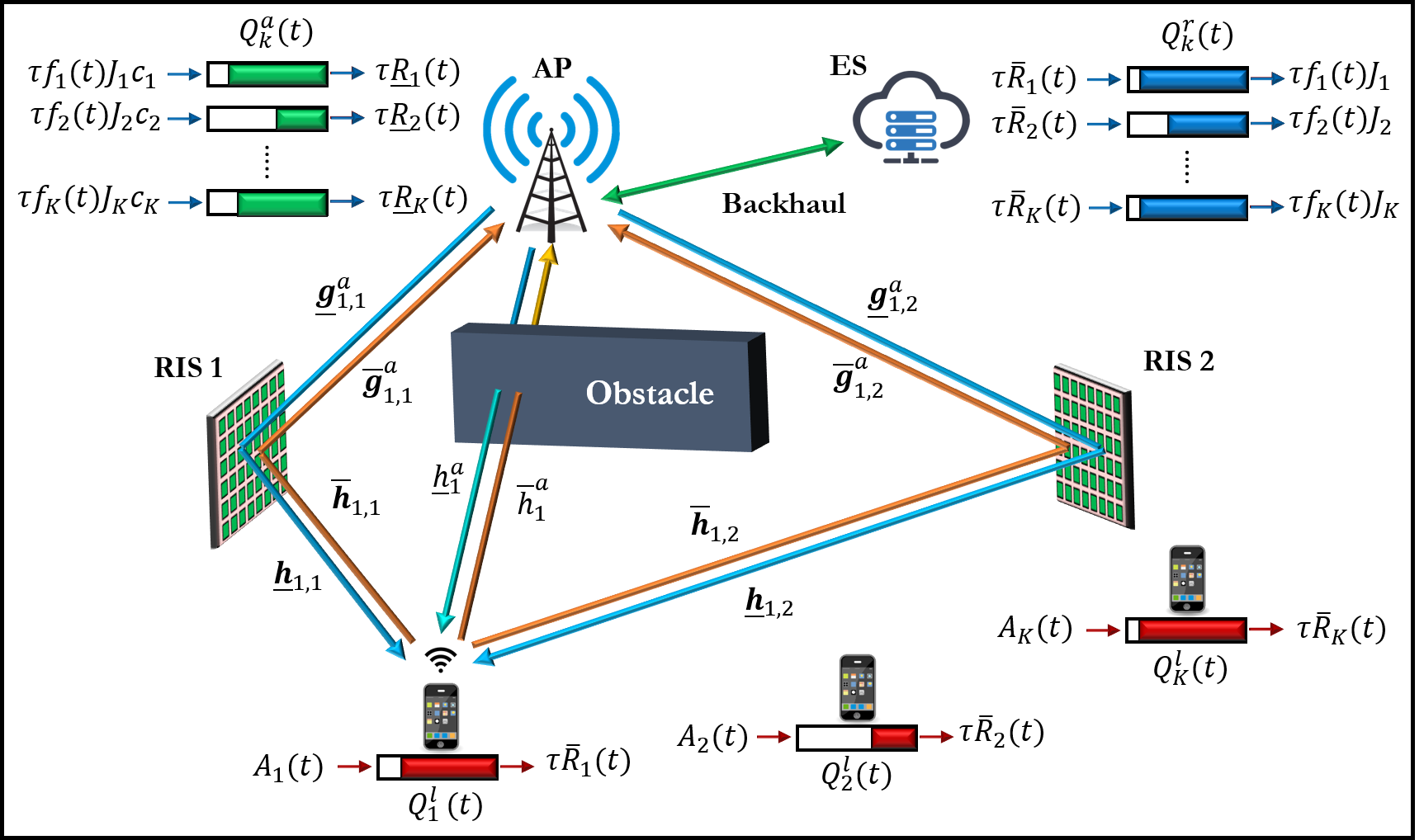}
    \caption{Network model}
    \label{fig:scenario}
\end{figure*}

\section{System model}

We consider a scenario with $K$ edge devices, an access point (AP) equipped with an edge server (ES), and $I$ RISs, as illustrated in Fig. \ref{fig:scenario}. Time is divided in slots indexed by $t$ and 
of equal duration $\tau_l$. We consider a block-fading model where the wireless channel is assumed to be static within each slot, whose duration $\tau_l$ is designed with respect to the channel coherence time. Also, the overall slot duration $\tau_l$ is divided into two portions: a period of $\tau_s$ seconds dedicated to control signaling, and a period of $\tau$ seconds for the actual three phases of computation offloading (i.e., uplink,  processing at the ES, and downlink). Here, we assume that control signaling happens before computation offloading due to the need of exchanging the state variables necessary to run the optimization algorithm and allocate radio and computation resources. Then, the total duration of the time slot is $\tau_l=\tau_s+\tau$. The quantification of $\tau$ and $\tau_s$ depends on typical trade-offs between complexity and performance. Indeed, a more accurate optimization could require a longer $\tau_s$, thus leaving less time for transmission and computation, and vice versa. We assume that the direct link between the users and the AP can be possibly impaired by the presence of obstacles, which attenuate or eventually block the communication, as shown in Fig.
\ref{fig:scenario}. The presence of the RISs helps counteract this detrimental effect by allowing alternative paths for communications between users and AP. However, also in the case without obstacles, the RISs typically enhance performance \cite{di2020smart}. We assume that synchronization of the system is enforced by the AP, which also controls the behavior of the RISs. In the sequel, we present the mathematical model of our dynamic system, considering RIS-enhanced communications, queueing model, and energy consumption.

\subsection{RIS-enhanced communications}

We consider a MEC system endowed with $I$ passive RISs, where the $i$-th RIS is composed of $N_i$ reflecting elements. The RIS $i$ at time $t$ is described by the reflectivity matrix:
\begin{equation}
    \boldsymbol{\Phi}_{i}(t)=\diag\{m_{i,1}(t) e^{j\phi_{i,1}(t)},\ldots,m_{i,N_i}(t) e^{j\phi_{i,N_i}(t)}\},
\end{equation}
for all $i,t$, where $m_{i,l}(t)\in\{0,1\}$ (i.e., the $l$-th reflective element of RIS $i$ is active or not at time $t$), and $\phi_{i,l}(t)\in\left\{ \frac{2k\pi}{2^{b_i}}\right\}_{k=0}^{2^{b_i}-1}$ (i.e., the phases are quantized using $b_i$ bits) \cite{mursia2020risma}. Equivalently, letting $v_{i,l}(t)=m_{i,l}(t) e^{j\phi_{i,l}(t)}$, we have
$\boldsymbol{\Phi}_{i}(t)=\diag\{\bv_{i}(t)\}$,  $\forall i,t$, where $\bv_{i}(t)=\{v_{i,l}(t)\}_{l=1}^{N_i}$, with \begin{equation}\label{phases}
v_{i,l}(t)\in \mathcal{S}_i= \left[0,\left\{e^{j\frac{2m\pi}{2^{b_i}}}\right\}_{m=0}^{2^{b_i}-1}\right], \quad \forall i,l,t.
\end{equation}
Also, let $\bv(t)=\{\bv_{i}(t)\}_{i=1}^I$. In the sequel, we will use the overline notation for uplink parameters, and the underline notation for downlink. Then, assuming a Single Input Single Output (SISO) communication system, the uplink transmission rate between user $k$ and the AP reads as:
\begin{align}\label{Uplink_rate}
    &\overline{R}_{k}(t)=\overline{B}_k\log_2\left(1+\overline{\alpha}_k(\bv(t)) \overline{p}_{k}(t)\right)
\end{align}
for all $k=1,\ldots,K$, where
\begin{align}\label{RIS_channel}
    &\displaystyle\overline{\alpha}_k(\bv(t))=\frac{\left|\overline{h}^a_{k}(t)+\displaystyle\sum _{i=1}^I \overline{\bh}_{k,i}(t)^T\,{\rm diag}(\bv_{i}(t))\,\overline{\bg}_{k,i}^a(t)\right|^2}{N_0 \overline{B}_k}
\end{align}
is the RIS-dependent normalized uplink channel coefficient, $\overline{p}_{k}(t)$ denotes the power transmitted by user $k$ at time $t$, and $\overline{h}^a_{k}(t)$ represents the direct uplink channel coefficient between user $k$ and the AP; whereas, $\overline{\bh}_{k,i}(t)\in \mathbb{C}^{N_i\times 1}$ and $\overline{\bg}_{k,i}^a(t)\in \mathbb{C}^{N_i\times 1}$ are vectors containing all the uplink channel coefficients between user $k$ and RIS elements, and between RIS elements and the AP, respectively. Specific models for $\overline{h}^a_{k}(t)$, $\overline{\bh}_{k,i}(t)$ and $\overline{\bg}_{k,i}^a(t)$ can be found in \cite{Basar20}. Furthermore, $\overline{B}_k$ denotes the bandwidth allocated to user $k$ for the uplink, and $N_0$ is the receiver noise power spectral density.

In our system, we assume that uplink and downlink happen simultaneously, using a frequency division duplexing scheme. Thus, similarly to (\ref{Uplink_rate}), the downlink transmission rate between the AP and user $k$ reads as:
\begin{align}\label{Downlink_rate}
    &\underline{R}_{k}(t)=\underline{B}_k\log_2\left(1+\underline{\alpha}_k(\bv(t)) \underline{p}_{k}(t)\right)
\end{align}
for all $k=1,\ldots,K$, where $\underline{p}_{k}(t)$ denotes the power transmitted by the AP toward user $k$ at time $t$, and
\begin{align}\label{RIS_channel_downlink}
    &\displaystyle\underline{\alpha}_k(\bv(t))=\frac{\left|\underline{h}^a_{k}(t)+\displaystyle\sum _{i=1}^I \underline{\bh}_{k,i}(t)^T\,{\rm diag}(\bv_{i}(t))\,\underline{\bg}_{k,i}^a(t)\right|^2}{N_0 \underline{B}_k}
\end{align}
is the RIS-dependent normalized downlink channel coefficient, where $\underline{h}^a_{k}(t)$, $\underline{\bh}_{k,i}(t)$,  $\underline{\bg}_{k,i}^a(t)$, and $\underline{B}_k$ are the downlink counterparts of the parameters in (\ref{RIS_channel}). Then, our goal is to optimize the uplink and downlink transmitted powers $\{\overline{p}_{k}(t)\}_{k=1}^K$ and $\{\underline{p}_{k}(t)\}_{k=1}^K$, respectively, jointly with the reflectivity parameters $\{\bv_{i}(t)\}_{i=1}^I$ of the available RISs.

\subsection{Evolution of data and computation queues}

We assume that each device $k$ generates $A_{k}(t)$ bits as input of the application to be executed at each time slot $t$. 
Then, a queueing system is used to model and control the dynamic data generation, transmission, and processing. In particular, at each time slot $t$, each user buffers data in a local queue $Q^l(t)$ and transmits them to the AP at the transmission rate $\overline{R}_{k}(t)$ (cf. \eqref{Uplink_rate}). Thus, the local queue update follows the rule:
\begin{equation}
\label{qup}
Q^l_{k}(t+1)=\max \left(0, Q^l_{k}(t)-\tau \overline{R}_{k}(t)\right)+A_{k}(t).
\end{equation}
The AP receives data from each device $k$ and sends the data to the ES, which processes $J_k$ bits-for-cycle, where $J_k$ is a parameter that depends on the application offloaded by device $k$. The ES provides a total CPU frequency $f_c(t)$ at each time slot, and a percentage $f_{k}(t)$ of it is allocated to process the task offloaded by user $k$ at slot $t$, such that $\sum_{k=1}^K f_{k}(t)\leq f_c(t)$ . Thus, the remote queue at the ES evolves as:
\begin{align}
\label{qcomp}
&Q^r_{k}(t+1)=\max \left(0, Q^r_{k}(t)-\tau f_{k}(t) J_k\right) \nonumber \\
&\qquad\qquad\qquad\qquad+\min\left(Q^l_{k}(t),\tau \overline{R}_{k}(t)\right),
\end{align}
where $\tau f_{k}(t) J_k$ are bits processed during each slot for UE $k$. Finally, the AP sends back to each user the bits resulting from the computation. Downlink communications can be incorporated considering an additional queue at the AP. To this aim, we assume that there is a linear dependence among the number of bits in input and those produced in output by the application running at the ES. Let denote by $c_k$ the ratio between output and input bits of the application required by user $k$. Thus, similarly to the previous models, the processed data are buffered in a queue $Q^a_{k}(t)$ at the AP and transmitted at a downlink rate $\underline{R}_{k}(t)$ (cf. (\ref{Downlink_rate})), with the update rule:
\begin{align}
\label{qdown}
&Q^a_{k}(t+1)=\max \left(0, Q^a_{k}(t) - \tau \underline{R}_{k}(t)\right)\nonumber \\
&\qquad\qquad\qquad\qquad
+c_k\cdot\min\left(Q^r_{k}(t),\tau f_{k}(t)J_k\right).
\end{align}
In this paper, we perform a joint optimization of RISs' phases, radio (uplink and downlink) and computation resources, considering the sum of communication and computation queues
\begin{equation}\label{sum_queues}
Q_{k}^{\textrm{tot}}(t)=Q_{k}^{l}(t)+Q_{k}^{r}(t)+Q_{k}^{a}(t)
\end{equation}
as a metric to quantify the overall delay experienced by data offloaded by each device. As we will show in the sequel, our aim is to keep the average value of $Q_{k}^{\textrm{tot}}(t)$ in (\ref{sum_queues}) (which is related to the average service delay through the Little's law \cite{little1961}) below a given threshold.

\subsection{Energy consumption}

In this paragraph, we evaluate the overall energy consumption of the system. In particular, we consider the energy spent for computation by the edge server, the energy spent for downlink communications by the AP, the energy spent for uplink transmission by each device and, finally, the energy spent by RISs to shape the wireless environment.

\textit{1) ES's energy consumption:} At the ES, the energy spent for computation is given by:
\begin{equation}\label{comp_power_serv}
    e_c(t)=\tau\gamma_c\left(f_c(t)\right)^3 +\tau_s\gamma_c f_m^3,
\end{equation}
where $f_c(t)$ and $\gamma_c$ are the CPU frequency and the effective switched capacitance of the ES processor, respectively \cite{Merl2021EML}. In \eqref{comp_power_serv}, we assumed (with the last term) that, during the portion $\tau_s$ of the slot, the server performs computations at speed $f_m$, which represents the minimum CPU frequency necessary to solve the optimization problem we will present in the sequel.

\textit{2) AP's energy consumption:} For the AP energy consumption, we exploit the concept from \cite{Debaille15,merluzzi2020d}, considering that a large portion of its energy is consumed only for being in \textit{active state} (i.e., to switch on RF chain, power amplifiers, power supply, analog front-end, digital baseband, and digital control). Then, we assume that the AP is able to enter low power sleep operation mode to save energy, a concept known as Discontinuous Transmission (DTX). In particular, let us denote by $p_a^{\rm on}$ the overall power consumption of the AP for being in active state. While in active state, the AP can transmit and/or receive. Instead, in sleep state, the AP can neither transmit nor receive. In active state, for downlink transmissions, the AP provides a maximum total power $P_a$ at each time slot, and a percentage $\underline{p}_{k}(t)$ of power is allocated for communicating with user $k$ at slot $t$, such that $\sum_{k=1}^K \underline{p}_{k}(t)\leq P_a$. In \cite{Debaille15}, four possible Sleep Modes (SM) are defined, with different minimum sleep periods, corresponding to the ODFM symbol, the sub-frame duration, the radio frame duration, and a standby mode. Here, we assume that the kind of sleep mode is selected a priori, while the choice of when being active or sleeping is performed online. To control the active and sleep state of the AP, we introduce the binary variable $I_a(t) \in \{0,1\}$, which is equal to $1$ if and only if the AP is in active state at time slot $t$. Also, in each time slot, the AP is forced to be active for the first $\tau_s$ seconds to perform Channel State Information (CSI) acquisition and control signaling, which we assume to be performed with the minimum power $\underline{P}_m$ required to achieve the target estimation and communication performance. The AP energy consumption at time slot $t$ is then given by:
\begin{align}\label{AP_energy}
&e_a(t)=\;\tau\Big(I_a(t)p_a^{\rm on}+I_a(t) \sum\nolimits_{k=1}^K \underline{p}_{k}(t)+(1-I_a(t))p_a^{\rm s}\Big)\nonumber \\
&\qquad\qquad\qquad\qquad
+\tau_s \left(p_a^{\textrm{on}}+\underline{P}_m\right),
\end{align}
where $p_a^{\rm s}$ represents the (low) power consumed in sleep mode.


\textit{3) UE's energy consumption:} Beyond uplink transmissions, we assume that each UE performs control signaling (during the first $\tau_s$ seconds) using the minimum power $\overline{P}_k$ needed to obtain a desired performance. Then, the energy spent by user $k$ at time $t$ is given by:
\begin{align}\label{user_energy}
    e_{k}(t)= \tau\, \overline{p}_{k}(t)I_a(t) +\tau_s \overline{P}_k,
\end{align}
for $k=1,\ldots,K$, where $\overline{p}_{k}(t)$ affects the uplink data rate as in \eqref{Uplink_rate}. Of course, from (\ref{user_energy}), if the AP is in sleep mode at time $t$, user $k$ does not spend energy for uplink transmission.

\textit{4) RIS's energy consumption:} The power consumption of a RIS depends on the type, the
resolution, and the number of its individual reflecting elements that effectively
perform phase shifting on the impinging signal \cite{ribeiro2018energy,mendez2016hybrid,huang2019reconfigurable}. In particular, let $p^{r}(b_i)$ be the power dissipated by each of the $N_i$ phase shifter of RIS $i$, assuming $b_i$-bit resolution. Typical power consumption values of each phase shifter are 1.5, 4.5, 6, and 7.8 mW for 3-, 4-, 5-, and 6-bit resolution phase shifting. In each time slot, each RIS is forced to have all active elements for the first $\tau_s$ seconds to perform CSI acquisition. Of course, if the AP is in sleep mode at time $t$, also the RIS is switched off. Then, the overall energy spent by RIS $i$ is given by:
\begin{align}\label{RIS_energy}
     e^{r}_i(t)= I_a(t)\tau p^{r}(b_i)\sum\nolimits_{l=1}^{N_i} |v_{i,l}(t)|^2+\tau_s N_i p^{r}(b_i),
 \end{align}
for $i=1,\ldots,I$, where we exploited the fact that each phase shift coefficient $v_{i,l}$ in (\ref{phases}) has either zero (if the $l$-th element is off) or unitary (if the $l$-th element is on) modulus. Thus, from (\ref{RIS_energy}), we can control the overall energy spent by the RISs at each time slot, acting on the number of active reflecting elements, and the value of the state variable $I_a(t)$.

In the following section, we will formulate the proposed dynamic strategy for RIS-empowered wireless network edge optimization, aimed at performing energy-efficient dynamic edge computing with guaranteed latency requirements.

\section{Dynamic RIS-empowered edge computing}

Our goal is to find the optimal scheduling of RISs' parameters (i.e., phase shifts), radio (i.e., powers, rates, AP/RISs/UEs duty cycles) and computation (i.e., CPU cycles) resources that minimizes the long-term average of a weighted sum of the energy consumption terms in (\ref{comp_power_serv})-(\ref{RIS_energy}), under constraints on the maximum average queue length in (\ref{sum_queues}). To this aim, we define the weighted sum energy as follows:
\begin{align}\label{weigthed_energy}
   \displaystyle e_{\sigma}^{\textrm{tot}}(t)=\sigma\sum_{k=1}^K e_{k}(t)+(1-\sigma)\left(e_c(t)+e_a(t)+\sum_{i=1}^I e^{r}_i(t)\right),
\end{align}
where $\sigma\in[0,1]$ is a weighting parameter to be chosen. For instance, choosing $\sigma=1$ leads to a pure \textit{user-centric} strategy; whereas, $\sigma=0$ determines a pure \textit{network-centric strategy}. An intermediate strategy, which we term as \textit{holistic}, can be obtained with $\sigma=0.5$. The use of this weighting parameter helps introduce more degrees of freedom and flexibility in the resource optimization, depending on the needs of the operators, users, and service providers. Using (\ref{weigthed_energy}), the problem can be mathematically cast as:
\begin{align}\label{Problem}
&\underset{\mathbf{\Psi}(t)} \min \;\;\;  \displaystyle \lim_{T\to\infty}\frac{1}{T}\sum_{t=1}^{T} {\mathbb{E}\left\{e_{\sigma}^{\textrm{tot}}(t)\right\}}\smallskip\\
&\quad\hbox{subject to}  \qquad \displaystyle (a) \;\;\lim_{T\to\infty}\,\frac{1}{T}\sum_{t=1}^T\mathbb{E}\left\{Q_{k}^{\textrm{tot}}(t)\right\}\leq Q_k^{\textrm{avg}},\;\; \forall k; \smallskip\nonumber\\
&\begin{rcases}
&\qquad (b)\quad \; I_a(t)\in\{0,1\} \;\quad \forall t; \smallskip;\\
&\qquad (c) \quad \;v_{i,l}(t)\in\mathcal{S}_i \qquad \forall i,l,t\smallskip;\\
&\qquad (d)\quad\; |v_{i,l}(t)|^2\leq I_a(t) \qquad \forall i,l,t; \smallskip;\\
&\qquad (e)\quad\; 0\leq \overline{p}_{k}(t)\leq P_k \,I_a(t), \qquad \forall k,t\smallskip;\\
&\qquad (f)\quad \;\displaystyle \underline{p}_{k}(t)\geq 0, \qquad \forall k,t;\smallskip;\\
&\qquad (g)\quad \;\displaystyle \sum_{k=1}^K \underline{p}_{k}(t)\leq P_a\, I_a(t), \;\quad \forall t;\smallskip\\
&\qquad (h)\quad \;\displaystyle f_{k}(t)\geq 0, \qquad \forall k,t;\smallskip\\
&\qquad (i)\quad \;\displaystyle \sum_{k=1}^K f_{k}(t)\leq f_c(t), \;\quad \forall t;\\
&\qquad (j)\quad \; f_c(t)\in\mathcal{F} \;\quad \forall t;
\end{rcases}
\mathcal{X}(t)\nonumber
\end{align}
where $\boldsymbol{\Psi}(t) = \big[\{\bv_{i}(t)\}_{i=1}^I, \{\overline{p}_{k}(t)\}_{k=1}^K,\{\underline{p}_{k}(t)\}_{k=1}^K,I_a(t),$ $\{f_{k}(t)\}_{k=1}^K, f_c(t)\big]$, and the expectations are taken with respect to the random channel states and data arrivals, whose statistics are supposed to be unknown. The constraints of \eqref{Problem} have the following meaning: $(a)$ the average queue lengths\footnote{More sophisticated constraints can also be imposed on the maximum tolerable delay \cite{Merluzzi2020URLLC}.} do not exceed a predefined value $Q_k^{\textrm{avg}}$ , for all $k$; $(b)$ The state variable $I_a(t)$ is binary; $(c)$ the RIS reflection coefficients take values from the discrete set $\mathcal{S}_i$ in (\ref{phases});
$(d)$ RIS modules can be active only if $I_a(t)=1$; $(e)$ the uplink transmission power is greater than zero and upper bounded by $P_k I_a(t)$, for all $k$; $(f)$ the downlink transmission power is greater than zero; $(g)$ the sum of all downlink transmitted powers is less than or equal to the maximum power $P_a$, or $0$ whenever the AP is inactive ($I_a(t)=0$); $(h)$ the CPU frequencies assigned to each device are greater than zero, and $(i)$ their sum is less than or equal to the ES CPU frequency $f_c(t)$; $(j)$ the ES CPU frequency takes values from a discrete set $\mathcal{F}$. Solving \eqref{Problem} is very challenging, because of the lack of knowledge of the statistics of the radio channels and task arrivals, and the inherent non-convexity. A further difficulty is related to the fact that the RISs are being optimized to handle, simultaneously, multiple data flows. Nevertheless, in the sequel, we will show how these problems can be effectively tackled resorting to stochastic Lyapunov optimization \cite{Neely10}, which enables low-complexity dynamic solutions for \eqref{Problem}.

\textit{Algorithmic solution: } We now convert the long-term optimization in \eqref{Problem} into a stability problem, hinging on stochastic Lyapunov optimization \cite{Neely10}. To deal with the long-term constraints in $(a)$, we introduce $K$ \textit{virtual queues} as:
\begin{equation}\label{Z}
    Z_{k}(t+1) = \max \Big \{0, Z_{k}(t) + \epsilon_k\left( Q^{\textrm{tot}}_{k}(t+1)-Q_k^{\textrm{avg}} \right) \Big \},
\end{equation}
$k=1,\ldots,K$, where $\{\epsilon_k\}_{k=1}^K$ are positive step sizes used to control the convergence speed of the algorithm. A virtual queue is a mathematical model that shows how the system is behaving in terms of constraint violations. Intuitively speaking, if a virtual queue grows too fast, the associated constraints are being violated and the system is not stable. Formally speaking, this translates into the \textit{mean rate stability} of the queues\footnote{A queue $X(t)$ is mean-rate stable if $\displaystyle\lim_{T\to \infty}\mathbb{E}\{X_T\}/T=0$.}, which is equivalent to satisfy the constraints $(a)$ in (\ref{Problem}) \cite{Neely10}. To this aim, we first define the Lyapunov function as
\begin{equation}\label{Lyapunov_fun}
    \mathcal{L}(t)=\mathcal{L}(\boldsymbol{\Theta}(t)) = \frac{1}{2} \sum_{k=1}^K Z^2_{k}(t),
\end{equation}
where $\boldsymbol{\Theta}(t)=\{Z_{k}(t)\}_{k=1}^K$, and then the \textit{drif-plus-penalty} function given by \cite{Neely10}:
\begin{equation}\label{Drift_pp}
    \Delta^p(t) = \mathbb{E}\Big\{\mathcal{L}(t+1)-\mathcal{L}(t) + V\cdot e_{\sigma}^{\textrm{tot}}(t)\big|\; \boldsymbol{\Theta}(t)\Big\}.
\end{equation}
The drift-plus-penalty function is the conditional expected change of $\mathcal{L}(t)$ over successive slots, with a penalty factor that weights the objective function of \eqref{Problem}, with a weighting parameter $V$. Now, if $\Delta_t^p$ is lower than a finite constant for all $t$, the virtual queues are stable and the optimal solution of \eqref{Problem} is asymptotically reached as $V$ increases \cite[Th. 4.8]{Neely10}. In practical scenarios with finite $V$ values, the higher is $V$, the more importance is given to the energy consumption, rather than to the virtual queue backlogs, thus pushing the solution toward optimality, while still guaranteeing the stability of the system. Thus, following similar arguments as in \cite{Neely10}, we proceed by minimizing an upper-bound of the drift-plus-penalty function in (\ref{Drift_pp}) in a stochastic fashion. After some simple algebra (similar as in \cite{merluzzi2020d} and omitted here due to space limitations), we obtain the following per-slot problem at each time $t$:
\begin{align}\label{inst_prob}
    &\min_{\boldsymbol{\Psi}(t)\in \mathcal{\widetilde{X}}(t)}\;\;  \sum\nolimits_{k=1}^K \Big[\big( Q_{k}^r(t)-Q^l_{k}(t)-Z_{k}(t)\big)\tau \overline{R}_{k}(t)\nonumber\\
    &\qquad\qquad    +\big(c_k Q_{k}^a(t)-Q^r_{k}(t)-Z_{k}(t)\big)\tau f_{k}(t)J_k \nonumber\\
    &\qquad\qquad-\big(Q^a_{k}(t)+Z_{k}(t)\big)\tau \underline{R}_{k}(t)\Big] +V \cdot  e_{\sigma}^{\rm tot}(t)
\end{align}
where $\mathcal{\widetilde{X}}(t)$ is the instantaneous feasible set, as defined in \eqref{Problem}, with the following modifications: i) constraint $(e)$ becomes $0\leq \overline{p}_k(t)\leq\widetilde{P}_k(t)I_a(t)$, where  $\widetilde{P}_{k}(t)=\min(P_k,\overline{P}_{k}(t))$, with $\overline{P}_{k}(t)$ denoting the minimum power needed to empty the local queue $Q^l_k(t)$ at time $t$; ii) constraint $(f)$ becomes $0\leq \underline{p}_k(t)\leq\underline{P}_k(t)I_a(t)$, where $\underline{P}_k(t)$ is the minimum power needed to empty the downlink queue $Q_k^a(t)$ at time $t$; iii) constraint $(h)$ becomes $0\leq f_{k}(t)\leq Q^r_{k}(t)/\tau J_{k}$. Because of the structure of set $\mathcal{\widetilde{X}}(t)$, \eqref{inst_prob} is a mixed-integer nonlinear optimization problem, which might be very complicated to solve. However, in the sequel, we will show how \eqref{inst_prob} can be split into sub-problems that admit low-complexity solution procedures for the optimal RIS parameters (i.e., the phase shifts of its elements), the uplink and downlink radio resources (i.e., powers, sleep mode and duty cycle), and the computation resources at the ES (i.e., CPU clock frequencies).

\section{Dynamic Radio Resource Allocation}\label{sec_resource_all}

The radio resource allocation problem aims at optimizing the AP duty cycle variable $I_a(t)$, the uplink and downlink transmission powers $\{\overline{p}_{k}(t)\}_{k=1}^K$, $\{\underline{p}_{k}(t)\}_{k=1}^K$, respectively, and the RIS reflectivity parameters  $\{\bv_{i}(t)\}_{i=1}^I$. From (\ref{Uplink_rate}) and (\ref{Downlink_rate}), it is clear that the presence of RISs couples uplink and downlink resource allocation, since transmission rates are affected by RISs in both directions. Then, from (\ref{inst_prob}), (\ref{weigthed_energy}), \eqref{Uplink_rate}, (\ref{Downlink_rate}), and (\ref{Problem}), the radio resource allocation problem reads as:
\begin{align}\label{radio_prob}
    &\min_{\boldsymbol{\Gamma}(t)}    \;\;-\sum_{k=1}^K\overline{U}_k(t)
    \log_2\left(1+\overline{\alpha}_k(\bv(t)) \overline{p}_{k}(t)\right)\\
    &\qquad\;
    -\sum_{k=1}^K\underline{U}_k(t)
    \log_2\left(1+\underline{\alpha}_k(\bv(t)) \underline{p}_{k}(t)\right) \nonumber\\
    &\qquad\;+ V \bigg[\sum_{k=1}^K\left( \sigma\tau\overline{p}_{k}(t)+\left(1-\sigma\right)\tau\underline{p}_{k}(t)\right)\nonumber\\
    &\qquad\;+(1-\sigma)I_a(t)\tau p_a^{\textrm{on}}+(1-\sigma)(1-I_a(t))\tau p_a^{\textrm{s}}\nonumber\\
    &\qquad\;+(1-\sigma)\tau\sum_{i=1}^I p^{r}(b_i)\sum_{l=1}^{N_i} |v_{i,l}(t)|^2 \bigg] \nonumber\\
&  \text{subject to} \quad  I_a(t)\in\{0,1\}; \quad 0\leq \overline{p}_{k}(t)\leq \widetilde{P}_{k}(t)I_a(t)\quad \forall k;\nonumber \\
&\qquad\qquad\quad v_{i,l}(t)\in\mathcal{S}_i, \quad |v_{i,l}(t)|^2\leq I_a(t) \;\;\forall i,l;\nonumber\\
&\qquad\qquad\quad 0\leq\underline{p}_{k}(t)\leq \underline{P}_{k}(t), \;\; \forall k;\quad\displaystyle \sum_{k=1}^K \underline{p}_{k}(t)\leq P_a \,I_a(t);\nonumber
\end{align}
where $\boldsymbol{\Gamma}(t) = [\bv(t), \{\overline{p}_{k}(t)\}_{k=1}^K,\{\underline{p}_{k}(t)\}_{k=1}^K,I_a(t)]$, and
\begin{align}
  &\overline{U}_k(t)=(Q_{k}^l(t)-Q^r_{k}(t)+Z_{k}(t)) \overline{B}_k\tau,  \label{U_up}\\
  &\underline{U}_k(t)=(Q_{k}^a(t)+Z_{k}(t)) \underline{B}_k\tau. \label{U_down}
\end{align}
Problem \eqref{radio_prob} is nonconvex due to the integer nature of the phase shifts and the active state variable of the AP (i.e., $I_a(t)$), and the coupling among variables induced by the presence of RISs. In principle, the global optimum solution of (\ref{radio_prob}) can be achieved through an exhaustive search over all the possible combinations of  $\{\bv_{i}(t)\}_{i=1}^I$ and $I_a(t)$, evaluating the optimal uplink and downlink powers, and selecting the set of variables that yields to the lowest value of the objective function in (\ref{radio_prob}). However, the complexity of this approach grows exponentially with the number $I$ of RISs, the maximum number $N=\max_i N_i$ of RIS elements, and the maximum cardinality $S=\max_i |\mathcal{S}_i|$ of the sets $\mathcal{S}_i$ in (\ref{phases}).
Since in the dynamic context considered in this paper resource allocation must take place in a very short amount of time, we follow an alternative (albeit simplified) optimization strategy. In particular, let us first notice that we can distinguish between two different cases.

\textit{\underline{Case $1$: $I_a(t)=0$}.} In this case, problem \eqref{radio_prob} is trivial, since the AP is in sleep state (thus not receiving and transmitting), and so are also the UEs and the RISs. Thus, the only feasible solution reads as:
\begin{align}\label{solution_sleep}
    &\overline{p}_k(t)=0, \;\;\forall k,\quad\underline{p}_k(t)=0, \;\;\forall k,\quad\bv_i(t)=0, \;\;\forall i.
\end{align}
In this case, the objective function of \eqref{radio_prob} boils down to:
\begin{equation}\label{objective_sleep}
    \Omega(I_a(t)=0)=V(1-\sigma)\tau p_a^{\textrm{s}}.
\end{equation}
The value in \eqref{objective_sleep} must be compared with the value of the objective function obtained in the following second case.

\textit{\underline{Case $2$: $I_a(t)=1$}.} In this case, the AP is available for transmission and/or reception, so that a solution is needed to select the uplink and downlink radio resources and the RIS reflectivity coefficients. In particular, problem \eqref{radio_prob} translates into the following simplified sub-problem:
\begin{align}\label{radio_prob_on}
    &\min_{\boldsymbol{\Psi}^r(t)}    \;\;-\sum_{k=1}^K\overline{U}_k(t)
    \log_2\left(1+\overline{\alpha}_k(\bv(t)) \overline{p}_{k}(t)\right)\\
    &\qquad\;-\sum_{k=1}^K\underline{U}_k(t)
    \log_2\left(1+\underline{\alpha}_k(\bv(t)) \underline{p}_{k}(t)\right)\nonumber\\
    &\qquad\;+ V \bigg[\sum_{k=1}^K\left( \sigma\tau\overline{p}_{k}(t)
    +\left(1-\sigma\right)\tau\underline{p}_{k}(t)\right)+(1-\sigma)\tau p_a^{\textrm{on}}  \nonumber\\
    &\qquad\; +(1-\sigma)\tau\sum_{i=1}^I p^{r}(b_i)\sum_{l=1}^{N_i} |v_{i,l}(t)|^2\bigg] \nonumber\\
& \quad\quad \text{subject to} \quad  0\leq \overline{p}_{k}(t)\leq \widetilde{P}_{k}(t)\;\;\forall k;\quad v_{i,l}(t)\in\mathcal{S}_i \;\;\forall i,l;\nonumber\\
&\qquad\qquad\qquad\quad 0\leq\underline{p}_{k}(t)\leq \underline{P}_{k}(t), \;\; \forall k;\quad\displaystyle \sum_{k=1}^K \underline{p}_{k}(t)\leq P_a.\nonumber
\end{align}
To solve \eqref{radio_prob_on}, we propose a greedy method that first optimizes (\ref{radio_prob_on}) with respect to the RIS reflectivity parameters $\{\bv_{i}(t)\}_{i=1}^I$, given the radio parameters, and then it selects the uplink and downlink powers. Indeed, given a fixed RISs configuration (i.e., for a given value of $\bv(t)$), \eqref{radio_prob_on} becomes strictly convex and decouples over uplink and downlink, admitting a simple closed form solution for $\{\overline{p}_{k}(t)\}_{k=1}^K$, and a water-filling like expression for $\{\underline{p}_{k}(t)\}_{k=1}^K$. The details of the three optimization steps (i.e., RISs, uplink, and downlink) are given next.

\subsection{RISs optimization}

To optimize \eqref{radio_prob} with respect to the RISs configuration, we notice that, for any value of $\overline{p}_{k}(t)$, if $\overline{U}_{k}(t)>0$, the $k$-th component of the first objective term in (\ref{radio_prob_on}) is minimized by increasing the normalized channel coefficients $\overline{\alpha}_{k}(\bv(t))$. A similar argument applies to the $k$-th component of the second objective term in (\ref{radio_prob_on}), which is minimized by increasing the normalized channel coefficient $\underline{\alpha}_{k}(\bv(t))$. Thus, letting $\mathcal{U}(t)=\{k\,|\,\overline{U}_{k}(t)>0\}$, 
we propose to exploit the following surrogate optimization function:
\begin{align}\label{Obj_greedy}
    \Delta^R(\bv(t))\,=\,&-\sum_{k\in\mathcal{U}(t)}\overline{U}_{k}(t)\overline{\alpha}_{k}(\bv(t))-\sum_{k=1}^K\underline{U}_{k}(t)\underline{\alpha}_{k}(\bv(t)) \nonumber\\
    &+ V (1-\sigma)\tau\sum_{i=1}^I p^{R}(b_i)\sum_{l=1}^{N_i} |v_{i,l}(t)|^2,
\end{align}
which represents a linear combination of the RIS energy term in (\ref{radio_prob_on}), weighted by the Lyapunov parameter $V$, and (negative) RIS-dependent uplink and downlink channel coefficients in (\ref{RIS_channel}) and (\ref{RIS_channel_downlink}),  weighted by the terms $\overline{U}_{k}(t)$ and $\underline{U}_{k}(t)$ in \eqref{U_up}-\eqref{U_down}, which depend on the communication, computing, and virtual queues' states. Intuitively, minimizing (\ref{Obj_greedy}), the RISs will be optimized to favor uplink and/or downlink communications (depending on the status of the cumulative parameters $\overline{U}_{k}(t)$ and $\underline{U}_{k}(t)$ for each user $k$), with a penalty on the energy spent for such improvement in communication performance. This is equivalent to a dynamic scheduling of the RIS resources to serve the users over uplink and/or downlink communications, depending on the status of the queues (i.e., $\overline{U}_{k}(t)$ and $\underline{U}_{k}(t)$) that quantify the system congestion. In other words, time plays the role of a further degree of freedom for the scheduling of the RISs, which are dynamically assigned by our procedure to serve uplink or downlink communications. Also, increasing the value of $V$, the minimization of (\ref{Obj_greedy}) leads to more sparse solutions for the vector $\bv(t)$, since it might be unnecessary to switch on all the reflecting elements to satisfy the average latency constraint in (\ref{Problem}). The steps of the proposed greedy method are illustrated in Algorithm \ref{alg:RIS}, which proceeds according to the following rationale. For each RIS $i$, the method greedily optimizes the reflectivity vector $\bar{\bv}_{i}$ (initialized at zero), iteratively selecting the coefficient $v_{i,l}\in\mathcal{S}_i$ that minimizes (\ref{Obj_greedy}), having fixed all the other parameters of RIS $i$ (i.e., $\bar{\bv}_{i,-l}$) and of the other RISs (i.e., $\bar{\bv}_{-i}$). This approach requires $O(S\overline{N})$ evaluations of (\ref{Obj_greedy}), with $\overline{N}=\sum_{i=1}^I N_i$, and leads to a non-increasing behavior of (\ref{Obj_greedy}) as more RIS reflecting elements are added and optimized. Interestingly, from (\ref{RIS_channel}) and (\ref{RIS_channel_downlink}),  we can also recast (\ref{Obj_greedy}) in the following compact matrix notation:
\begin{align}\label{Obj_greedy_compact}
    \Delta^R(\bv(t))\,=\, \widetilde{\bv}(t)^H \mH(t) \widetilde{\bv}(t)
\end{align}
where $\widetilde{\bv}(t)=[1,\bv_1(t)^T,\ldots,\bv_I^T]^T\in \mathbb{C}^{(\overline{N}+1)\times 1}$, and the matrix $\mH(t)\in \mathbb{C}^{(\overline{N}+1)\times (\overline{N}+1)}$ is built as
\begin{align}\label{matrix_H}
    &\mH(t)=\overline{\mH}(t)+\underline{\mH}(t)+V\cdot \mD,
\end{align}
where we used the following definitions:
\begin{align}
    &\overline{\mH}(t)=- \sum_{k\in\mathcal{U}(t)} \overline{U}_k(t) \overline{\bz}^*_k(t) \overline{\bz}^T_k(t), \nonumber\\ &\overline{\bz}_k(t)=\overline{\bh}_k(t) \odot \overline{\bg}^a_k(t), \nonumber\\
    & \overline{\bh}_k(t)=[\overline{h}^a_k(t),\overline{\bh}_{k,1}(t)^T,\ldots,\overline{\bh}_{k,I}(t)^T]^T ,\nonumber\\
    &\overline{\bg}^a_k(t)=[1,\overline{\bg}^a_{k,1}(t)^T,\ldots,\overline{\bg}^a_{k,I}(t)^T]^T,    \nonumber\\
    &\underline{\mH}(t)=- \sum_{k=1}^K \underline{U}_k(t) \underline{\bz}^*_k(t) \underline{\bz}^T_k(t),\nonumber\\
    &\underline{\bz}_k(t)=\underline{\bh}_k(t) \odot \underline{\bg}^a_k(t), \nonumber
\end{align}
\begin{align}
    & \underline{\bh}_k(t)=[\underline{h}^a_k(t),\underline{\bh}_{k,1}(t)^T,\ldots,\underline{\bh}_{k,I}(t)^T]^T, \nonumber\\
    &\underline{\bg}^a_k(t)=[1,\underline{\bg}^a_{k,1}(t)^T,\ldots,\underline{\bg}^a_{k,I}(t)^T]^T, \nonumber\\
    & \mD = {\rm diag}\left\{0,p^{r}(b_1)\mathbf{1}_{N_1}^T,\ldots,p^{r}(b_I)\mathbf{1}_{N_I}^T\right\}, \nonumber
\end{align}
with $\odot$ denoting the Hadamard product. Since, the computation of (\ref{Obj_greedy_compact}) requires $O\left(\left(\overline{N}+1\right)^2\right)$ operations, the complexity of the greedy procedure in Algorithm 1 is given by $O\left(S\overline{N}^3\right)$. Of course, once put in the form (\ref{Obj_greedy_compact}), the calculation of the surrogate can exploit very efficient algorithms for (sparse) matrix-vector multiplications.

\begin{algorithm}[t]
\textbf{Input:} $V$, $\{p^r(b_i)\}_{i=1}^I$, $\{\overline{U}_k(t)\}_{k\in\mathcal{U}(t)}$, $\{\underline{U}_k(t)\}_{k=1}^K$, $\{\overline{\bh}_k(t)\}_{k=1}^K$, $\{\overline{\bg}_k^a(t)\}_{k=1}^K$, $\{\underline{\bh}_k(t)\}_{k=1}^K$, $\{\underline{\bg}_k^a(t)\}_{k=1}^K$. \smallskip

Set $\bar{\bv}_{i}=\mathbf{0}$ $\forall i$, and evaluate matrix $\mH(t)$ in \eqref{matrix_H}.\smallskip\\
\For{$i=1:I$\smallskip}
{
\For{$l=1:N_i$\smallskip}
{
$\bar{v}_{i,l}=\displaystyle\arg\min_{v_{i,l}\in\mathcal{S}_i}\;\Delta^R(v_{i,l};\bar{\bv}_{i,-l},\bar{\bv}_{-i})$
}
Set $\bv_{i}(t)=[\bar{v}_{i,1},\cdots,\bar{v}_{i,N_i}]^T$\smallskip\\
}

\textbf{Output:} $\{\bv_{i}(t)\}_{i=1}^I$
\caption{\textbf{: Greedy RIS optimization}}
\label{alg:RIS}
\end{algorithm}


\textit{Remark 1:} Of course, there are no guarantees that the proposed RISs optimization procedure (i.e., Algorithm 1) finds the global optimal solution of (\ref{radio_prob_on}) with respect to the RISs configuration, since it represents only an approximation of it (but attainable with low-complexity). However, in the context of stochastic Lyapunov optimization, our approach can be interpreted as a $C$-additive approximation \cite[p. 59]{Neely10}, which admits inexact solutions (with bounded error) of the drift-plus-penalty method in (\ref{inst_prob}) at each time $t$. Indeed, for any given value of the (real and virtual) queues at time $t$, the objective and the feasible set of (\ref{inst_prob}) are bounded for all $t$, and thus the (expected) difference of the objective values achieved by an exhaustive search procedure (striking the optimum) and the proposed greedy approach is always upper-bounded by a finite constant $C$. In Sec. IV, we will numerically assess the performance of the proposed resource allocation strategy.

\textit{Block optimization of RISs:} Even if the complexity of the greedy procedure in Algorithm 1 is sufficiently low, in practical scenarios one might still desire an even faster procedure. To this aim, we might divide the $N_i$ modules of RIS $i$ in $N_b$ blocks, where the elements of each block are phase-shifted in the same way. Then, proceeding as in Algorithm 1, each block of RIS $i$ is greedily optimized selecting the phase shift coefficient (equal for each element of the block) that leads to the largest decrease of the surrogate objective in (\ref{Obj_greedy_compact}). Assuming for simplicity that the number of blocks is the same for all RISs, the complexity of Algorithm 1 scales as $O\left(S I N_b\overline{N}^2\right)$. With respect to the full optimization, the complexity is reduced of a factor $\overline{N}/I N_b$, which is of course paid in terms of an overall reduction of performance. This complexity-performance trade-off will be assessed through numerical results in Sec. \ref{sec:Numerical_results}.

\subsection{Uplink radio resource allocation} Once the RIS configuration $\bv(t)$ has been fixed, from (\ref{radio_prob_on}), the uplink radio resource allocation decouples from downlink, and reads as:
\begin{align}\label{uplink_radio_prob}
    \hspace{-.1cm}\min_{\{\overline{p}_{k}(t)\}_{k=1}^K}    \hspace{-.2cm}&-\sum_{k=1}^K\overline{U}_k(t)
    \log_2\left(1+\overline{\alpha}_k(\bv(t)) \overline{p}_{k}(t)\right) + V\sigma\tau \sum_{k=1}^K \overline{p}_{k}(t) \nonumber\\
& \quad\quad \text{subject to} \quad \quad 0\leq \overline{p}_{k}(t)\leq \widetilde{P}_{k}(t),\quad \forall k.
\end{align}
Problem (\ref{uplink_radio_prob}) is convex, with an additive strictly convex objective that decouples over the users. Now, imposing the Karush-Kuhn-Tucker (KKT) conditions of (\ref{uplink_radio_prob}), it is easy to see that the problem admits a closed form solution for the optimal $\{ \overline{p}_{k}(t)\}_{k=1}^K$. In particular, the set $\mathcal{U}(t)=\{k\,|\,\overline{U}_{k}(t)>0\}$ previously used in (\ref{Obj_greedy}) takes the role of the set of transmitting users. Indeed, from a rapid inspection of (\ref{uplink_radio_prob}), it is clear that user $k$ does not transmit (i.e., $\overline{p}_{k}(t)=0$) if $\overline{U}_{k}(t)<0$ (since both terms of the objective function in \eqref{uplink_radio_prob} are monotone non decreasing functions of $\overline{p}_{k}(t)$). Thus, we obtain the simple closed form solution for the optimal uplink powers:
\begin{align}\label{Uplink_powers}
\overline{p}_{k}(t)=\begin{cases}
    & \hspace{-.3cm} \displaystyle\left[\frac{\overline{U}_{k}(t)}{V\tau\log 2}-\frac{1}{\overline{\alpha}_k(\bv(t))}\right]_0^{\widetilde{P}_{k}(t)}, \;\;\hbox{if $k\in\mathcal{U}_t$;}\medskip\\
    & \hspace{-.2cm}0\,, \qquad\qquad\hspace{2.18cm}\;\;\; \hbox{if  $k\notin\mathcal{U}_t$.}
    \end{cases}
\end{align}
As expected, for all $k$, the transmission powers at time $t$ in (\ref{Uplink_powers}) are affected by the RIS-dependent uplink channel coefficient $\overline{\alpha}_k(\bv(t))$, and the status of the communication, computation, and virtual queues embedded into $\overline{U}_{k}(t)$ (cf. (\ref{U_up})).

\subsection{Downlink radio resource allocation}

Once the RISs configuration $\bv(t)$ has been fixed, the downlink radio resource allocation problem optimizes the downlink transmission powers $\{\underline{p}_{k}(t)\}_{k=1}^K$. From (\ref{radio_prob_on}), we obtain:
\begin{align}\label{downlink_radio_prob}
     \min_{\{\underline{p}_{k}(t)\}_{k=1}^K}   &-\sum_{k=1}^K\underline{U}_k(t)
   \log_2\left(1+\underline{\alpha}_k(\bv(t)) \underline{p}_{k}(t)\right) \nonumber\\
   & \quad\quad+ V (1-\sigma)\tau\left(\sum_{k=1}^K\underline{p}_k(t)+p_a^{\textrm{on}}\right)\\
& \hspace{-.7cm}\text{subject to}\quad 0\leq\underline{p}_{k}(t)\leq \underline{P}_{k}(t), \;\;\forall k;\quad \displaystyle \sum_{k=1}^K \underline{p}_{k}(t)\leq P_a.\nonumber
\end{align}
Problem \eqref{downlink_radio_prob} is convex, and its solution can be found very efficiently imposing the KKT conditions. In particular, let us write the Lagrangian associated to  \eqref{downlink_radio_prob}, which reads as:
\begin{align}
    &\mathcal{L}=-\sum_{k=1}^K\underline{U}_k(t)
   \log_2\left(1+\underline{\alpha}_k(\bv(t)) \underline{p}_{k}(t)\right) \nonumber\\
   &+ V(1-\sigma)\tau \left(\sum_{k=1}^K\underline{p}_k(t)+p_a^{\textrm{on}}\right)-\sum_{k=1}^K\beta_k\underline{p}_k(t)\nonumber\\
    &+\sum\nolimits_{k=1}^K \gamma_k(\underline{p}_k(t)-\underline{P}_k(t))+\nu\left(\sum\nolimits_{k=1}^K\underline{p}_k(t)-P_a\right)
\end{align}
\begin{algorithm}[t]
\textbf{Input:} $V$,  $\{\underline{\alpha}_k(t)\}_{k=1}^K$, $\{\underline{U}_k(t)\}_{k=1}^K$. \smallskip

Let $\{\underline{p}_k^c(t)\}_{k=1}^K$ be the candidate powers obtained as \eqref{downlink_powers1}.\smallskip\\
\eIf{$\sum_{k=1}^K\underline{p}_ k^c(t)\leq p_a^{\max}$}{\smallskip Set $\underline{p}_k(t)=p_k^c(t)$ for all $k$}{
Compute the optimal value of $\underline{p}_k(t)$ as in \eqref{downlink_powers}}

\textbf{Output:} $\{\underline{p}_{k}(t)\}_{k=1}^K$
\caption{\textbf{: Downlink Radio Resource Allocation}}
\label{alg:downlink}
\end{algorithm}
Then, the KKT conditions are given by:
\begin{align}\label{KKT}
    &i)\;-\frac{\underline{U}_k(t)\underline{\alpha}_k(t)}{\log(2)\left(1+\underline{\alpha}_k(t)\underline{p}_k(t)\right)}\nonumber\\
    &\qquad\qquad\quad+V(1-\sigma)\tau-\beta_k+\gamma_k+\nu=0;\\
    &ii)\;\beta_k\geq 0;\quad\underline{p}_k(t)\geq 0;\quad \beta_k\underline{p}_k(t)=0,\quad \forall k;\nonumber\\
    &iii)\;\gamma_k\geq 0;\quad\underline{p}_k(t)\leq \underline{P}_k(t);\quad \gamma_k\left(\underline{p}_k(t)-\underline{P}_k(t)\!\right)=0,\; \forall k;\nonumber\\
    &iv)\;\nu\geq 0;\;\; \sum\nolimits_{k=1}^K\underline{p}_k(t)\leq P_a;\;\;\nu\left(\sum\nolimits_{k=1}^K\underline{p}_k(t)-P_a\right)=0. \nonumber
\end{align}
Now, let us consider two cases. First of all, if we assume that $\sum_{k=1}^K\underline{p}_k(t)<P_a$, we have $\nu=0$ due to condition $iv)$ in \eqref{KKT}. Then, from condition $i)$, the optimal solution is:
\begin{align}\label{downlink_powers1}
    \underline{p}_{k}(t)=\left[\frac{\underline{U}_k(t)}{V(1-\sigma)\log 2}-\frac{1}{\underline{\alpha}_k(\bv(t))}  \right]_0^{\underline{P}_k(t)} \;\;\forall k.
\end{align}
This means that, evaluating \eqref{downlink_powers1} for all $k$, if $\sum_{k=1}^K\underline{p}_k(t)\leq P_a$, then \eqref{downlink_powers1} is also the global optimal solution of \eqref{downlink_radio_prob}, since it satisfies all the KKT conditions. In the second case, given \eqref{downlink_powers1}, if $\sum_{k=1}^K\underline{p}_k(t)> P_a$, we must have $\nu>0$, and the optimal solution of \eqref{downlink_radio_prob} is found by imposing $\sum_{k=1}^K\underline{p}_k(t)= P_a$ due to condition $iv)$ in \eqref{KKT}. In this case, from condition $i)$ in \eqref{KKT}, the solution of \eqref{downlink_radio_prob} admits a water-filling like structure
\cite{palomar2005practical} (whose practical implementation requires at most $K$ iterations). More specifically, the optimal poweres read as:
\begin{align}\label{downlink_powers}
    \underline{p}_{k}(t)=\left[\frac{\underline{U}_k(t)}{[V(1-\sigma)+\nu]\log 2}-\frac{1}{\underline{\alpha}_k(\bv(t))}  \right]_0^{\underline{P}_k(t)} \;\;\forall k,
\end{align}
where $\nu$ is the Lagrange multiplier chosen to satisfy the power budget constraint with equality, i.e., $\sum_{k=1}^K \underline{p}_{k}(t)= P_a$.
The overall procedure is summarized in Algorithm \ref{alg:downlink}, and is very efficient. Indeed, in the case the closed-form solution in \eqref{downlink_powers1} is such that $\sum_{k=1}^K \underline{p}_{k}(t)\leq P_a$, the procedure stops and the water-filling solution in (\ref{downlink_powers}) is not needed.


\textit{Overall procedure for radio resource allocation:} Using Algorithm 1, (\ref{Uplink_powers}), and Algorithm 2, we have the proposed solution to problem \eqref{radio_prob_on}, i.e., the solution of problem \eqref{radio_prob} when the AP is active, i.e., $I_a(t)=1$. Now, to decide  the AP state variable $I_a(t)$, we need to compare the value of the objective function of \eqref{radio_prob} in the active case with the one achieved in the sleep state, i.e., (\ref{objective_sleep}). Then, denoting by $\bv^{\textrm{on}}(t)$, $\{\underline{p}_k^{\textrm{on}}\}_{k=1}^K$, and $\{\overline{p}_k^{\textrm{on}}\}_{k=1}^K$ the solution obtained with $I_a(t)=1$ (through Algorithm 1, \eqref{Uplink_powers}, and Algorithm 2), the value of the objective function of \eqref{radio_prob} reads as:
\begin{align}\label{objective_on}
    &\Omega(I_a(t)=1)=-\sum_{k=1}^K\overline{U}_k(t)
    \log_2\left(1+\overline{\alpha}_k(\bv^{\textrm{on}}(t)) \overline{p}_{k}^{\textrm{on}}(t)\right)\nonumber\\
    &\quad-\sum_{k=1}^K\underline{U}_k(t)
    \log_2\left(1+\underline{\alpha}_k(\bv^{\textrm{on}}(t)) \underline{p}_{k}^{\textrm{on}}(t)\right) \nonumber\\
    &\quad+ V\tau \Bigg[\sum_{k=1}^K\left( \sigma\overline{p}_{k}^{\textrm{on}}(t)+(1-\sigma)\underline{p}_{k}^{\textrm{on}}(t)\right)\nonumber\\
    &\quad+ (1-\sigma)\!\!\left(\! p_a^{\textrm{on}}+\sum_{i=1}^I p^{R}(b_i)\sum_{l=1}^{N_i} |v_{i,l}^{\textrm{on}}(t)|^2\right)\Bigg].
\end{align}
Then, the final solution of \eqref{radio_prob} is found by comparing \eqref{objective_sleep} and \eqref{objective_on}. Indeed, if $O(I_a(t)=0)\leq O(I_a(t)=1)$, the solution is given by \eqref{solution_sleep}. Otherwise, the solution is given by Algorithm 1, \eqref{Uplink_powers}, and Algorithm 2. The overall procedure for dynamic radio resource allocation is described in Algorithm \ref{alg:rra}.

\begin{algorithm}[t]
\textbf{Input:} All the inputs of Algorithms 1 and 2. \smallskip\\
Compute the objective $\Omega_0$ as in \eqref{objective_sleep}.\smallskip\\
Evaluate:
\begin{itemize}
    \item $\bv_i^{\textrm{on}}(t),\forall i$, with Algorithm \ref{alg:RIS};
    \item $\overline{p}_k^{\textrm{on}}(t), \forall k$, as in \eqref{Uplink_powers};
    \item $\underline{p}_k^{\textrm{on}}(t), \forall k$, with Algorithm \ref{alg:downlink}.\smallskip
\end{itemize}
Compute the objective $\Omega_1$ as in \eqref{objective_on}. \smallskip\\
\eIf{$\Omega_0\leq \Omega_1$\smallskip}{$I_a(t)=0$; $\;\;\overline{p}_k(t)=\underline{p}_k(t)=0,\forall k$; $\;\;\bv_i=0,\forall i$;}{\smallskip$I_a(t)=1$; $\;\;\overline{p}_k(t)=\overline{p}_k^{\textrm{on}}(t)$, $\;\;\underline{p}_k(t)=\underline{p}_k^{\textrm{on}}(t),\forall k$; \\$\bv_i(t)=\bv_i^{\textrm{on}}(t),\forall i$;\smallskip}
\textbf{Output:} $I_a(t)$, $\{\overline{p}_{k}(t)\}_{k=1}^K$, $\{\underline{p}_{k}(t)\}_{k=1}^K$, $\{\bv_i(t)\}_{i=1}^I$
\caption{\textbf{: Dynamic Radio Resource Allocation}}
\label{alg:rra}
\end{algorithm}

\section{Dynamic Allocation of Computing Resources}

The computing resource allocation problem optimizes the CPU frequencies $\{f_{k}(t)\}_{k=1}^K$ assigned by the server to the devices, and the overall ES frequency $f_c(t)$. From (\ref{inst_prob}), letting $Y_{k}(t)=(-c_kQ_k^a(t)+Q_{k}^r(t)+Z_{k}(t))J_k\tau$, we obtain
\begin{align}\label{slot_opt_cpu}
&\min_{\{f_{k}(t)\}_{k=1}^K,\,f_c(t)} -\sum_{k=1}^K Y_{k}(t)  f_{k}(t) + V(1-\sigma)\tau \gamma_s(f_c(t))^3\nonumber\\
& \quad\quad \text{subject to} \quad 0\leq f_{k}(t)\leq \frac{Q^r_{k}(t)}{\tau J_{k}},\;\;\forall k;\\
&\qquad\qquad\qquad\quad\displaystyle \sum_{k=1}^K f_{k}(t)\leq f_c(t); \qquad f_c(t)\in\mathcal{F}.\nonumber
\end{align}
The CPU frequency $f_c(t)$ in (\ref{slot_opt_cpu}) is assumed to belong to a fixed discrete set $\mathcal{F}$. Thus, for a given $f_c(t)\in\mathcal{F}$, problem \eqref{slot_opt_cpu} is linear in $\{f_{k}(t)\}_{k=1}^K$, and the optimal frequencies can be achieved using the simple procedure in Algorithm \ref{alg:cpu}. Intuitively, Algorithm 4 assigns the largest portions of $f_c(t)$ to the devices with largest values of $Y_{k}(t)$, and requires at most $K$ steps. Also, letting $\mathcal{C}(t)=\{k\,|\, Y_{k}(t)>0\}$, it is clear that the ES assigns a non-zero CPU frequency only to the devices belonging to $\mathcal{C}(t)$. Then, letting $\{f_{k}(f_c(t))\}_{k=1}^K$ be the optimal frequencies assigned at the users for a given $f_c(t)\in\mathcal{F}$ (using Algorithm 4), the optimal ES frequency $f_c(t)$ is given by:
\begin{equation}\label{optimal_ES_frequency}
    f_c(t)= \arg\min_{f_c\in\mathcal{F}} \;\;-\sum_{k\in\mathcal{C}(t)} Y_{k}(t) f_{k}(f_c)+ V (1-\sigma)\tau\gamma_s(f_c)^3.
\end{equation}
Finally, the variables $f_c(t)$ and $\{f_{k}(f_c(t))\}_{k=1}^K$ represent the global optimal solution of \eqref{slot_opt_cpu} at time $t$. The worst case number of scalar operations needed by this procedure is $O(K|\mathcal{F}|)$, which is affordable in many practical scenarios.

\begin{algorithm}[t]
\textbf{Input:} $\{Y_{k}(t)\}_k$,  $\mathcal{C}(t)$, $\{Q^r_{k}(t)\}_k$, $\{J_k\}_k$, $f_c$, $K$.\smallskip\\
Set $f_{\rm av}=f_c$, $\{f_k\}_{k=1}^K=0$, and $\mathcal{C}=\mathcal{C}(t)$\smallskip

\While{$f_{\rm av}>0$\smallskip}
{Find $\widetilde{k}=\displaystyle\arg \max_{k\in\mathcal{C}} \;Y_{k}(t)$\smallskip\\
Set $f_{\widetilde{k}}=\min\left(\displaystyle\frac{Q^r_{\widetilde{k}}(t)}{\tau J_{\widetilde{k}}},f_{\rm av}\right)$\smallskip\\
Set $\mathcal{C}=\mathcal{C}-\left\{\widetilde{k}\right\}$; $f_{\rm av}=f_{\rm av}-f_{\widetilde{k}}$\medskip\\
If $\mathcal{C}=\emptyset$ $\rightarrow$ break\smallskip}
\textbf{Output:} $\{f_k\}_{k=1}^K$
\caption{\textbf{: Optimal scheduling of CPU frequencies}
\label{alg:cpu}}
\end{algorithm}

Finally, the overall procedure for the proposed resource allocation strategy for RIS-empowered dynamic mobile edge computing is summarized in Algorithm \ref{alg:RIS-MEC}. The method is fully dynamic and optimizes variables on-the-fly via closed form expressions or low-complexity procedures (which do not require asymptotic convergence of iterative algorithms), based on instantaneous realization of the involved random variables (i.e., wireless channels, and data arrivals). In the next section, we assess the performance of the proposed dynamic optimization strategy via numerical simulations.

\begin{algorithm}[t]
\SetAlgoLined
\vspace{.1cm}
Set the Lyapunov trade-off parameter $V$, $Z_k(0)$, $\epsilon_k$, for all $k$. In each time slot $t\geq 0$, repeat the following steps:

\begin{enumerate}
    \item Find the RISs phase shifts $\{\bv_i\}_{i=1}^I$ and the radio parameters $I_a(t)$, $\{\overline{p}_{k}(t)\}_{k=1}^K$, $\{\underline{p}_{k}(t)\}_{k=1}^K$ using Algorithm \ref{alg:rra};\smallskip
   \item Solve the CPU scheduling problem in (\ref{optimal_ES_frequency}), where the optimal frequencies $\{f_k(t)\}_{k=1}^K$ assigned by the edge server are given by Algorithm \ref{alg:cpu}. \smallskip
   \item Perform the mobile edge computing task. \smallskip
   \item Update the physical queues as in \eqref{qup}, \eqref{qcomp},  \eqref{qdown}, and the virtual queues as in \eqref{Z}.
\end{enumerate}
\caption{\textbf{: RIS-empowered dynamic MEC}}
\label{alg:RIS-MEC}
\end{algorithm}

\section{Numerical results}\label{sec:Numerical_results}

\textit{Simulation setup:} We consider a scenario similar to Fig. \ref{fig:scenario}, with $K=5$ users wishing to offload their applications to the ES, through a wireless connection with an AP operating at $f_0=28$ GHz. The total available bandwidth is $B=100$ MHz, equally shared among users, and a noise power spectral density $N_0=-174$ dBm/Hz. At each time slot, the SISO channels $\{\underline{h}_k^a\}_{k=1}^K$, $\{\overline{h}_k^a\}_{k=1}^K$ between the users and the AP, the channels $\{\underline{\bh}_{k,i}\}_{i,k}$, $\{\overline{\bh}_{k,i}\}_{i,k}$ between the users and RISs, and the channels $\{\underline{\bg}^a_{k,i}\}_{i,k}$, $\{\overline{\bg}^a_{k,i}\}_{i,k}$ between the RISs and the AP are generated through the available tool SimRIS \cite{Basar20}. In particular, denoting by $(x,y,z)$ the 3D coordinates of an element, we set the following positions and parameters:
\begin{itemize}
    \item 1 AP at $(0,25,2)$, with  $p_a^{\textrm{on}}=2.2$ W, $p_a^{\textrm{s}}=278$ mW and $P_a=24$ dBm, according to a pico-cell case \cite{Debaille15}.
    \item 2 RISs with $N_i=100$ elements at ($33,28,2$) and ($33,18,2$). We assume that each phase can be encoded with $b_i$ bits (cf. \eqref{phases}), with $b_i$ ranging from $1$ to $3$ across the different simulations. Therefore, the energy consumption assumed for controlling a single element is set to (cf. \eqref{RIS_energy}) $p^R(b_i=1)=0.5$ mW, $p^R(b_i=2)=1$ mW, and $p^R(b_i=3)=1.5$ mW.
    \item $5$ users at ($34,20,1$), ($35,20,1$), ($36,20,1$), ($36,22,1$), and ($35,22,1$). The maximum transmit power of a generic user $k$ is set to $P_k=100$ mW.
\end{itemize}
All channels experience a coherence time equal to the total slot duration $\tau_l=10$ ms. A portion $\tau_s=1$ ms is devoted to control signaling and optimization. Thus, the queues are drained for $\tau=9$ ms, while the arrival rate is computed as $\bar{A}_k=\mathbb{E}\{A_k(t)/\tau_l\}$ and is set to $100$ kbps with Poisson distribution, for all users. As depicted in Fig. \ref{fig:scenario}, we assume that an obstacle obscures the direct communication between the users and the AP with $30$ dB of additional path loss. From an application point of view, we consider a conversion factor $J_k=10^{-3}, \forall k$ (cf. \eqref{qcomp}), and $c_k=1,\forall k$ (cf. \eqref{qdown}). An average constraint on the E2E delay equal to $50$ ms is imposed (cf. (\ref{Problem})). Also, we assume that the ES frequency $f_t^s$ (cf. \eqref{comp_power_serv}) can be selected in the finite set $\mathcal{F}=[0,0.01,0.02,\ldots,1]\times f_{\max}$, with $f_{\max}=4.5$ GHz, while the effective switched capacitance of the processor is set to $\gamma_s=10^{-27} \; W\cdot s^3$ .

\begin{figure}[t]
    \centering
    \includegraphics[width=7 cm]{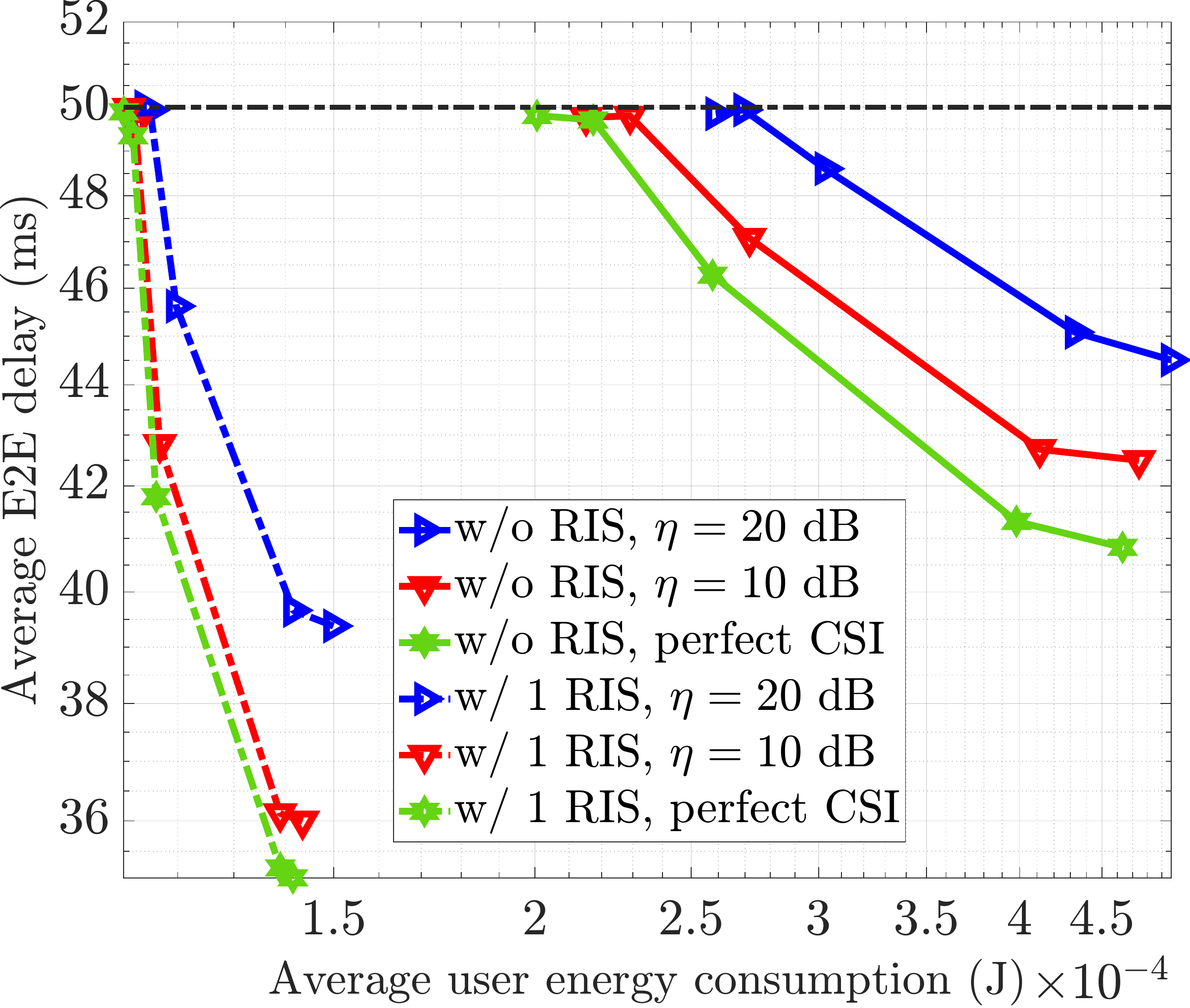}
    \caption{Avg. delay vs. user energy (single user)}
    \label{fig:tradeoff_su}
\end{figure}

\begin{figure*}[t]
    \centering

    \subfloat[Avg. delay vs. average system energy]{
        \includegraphics[width=0.32\textwidth]{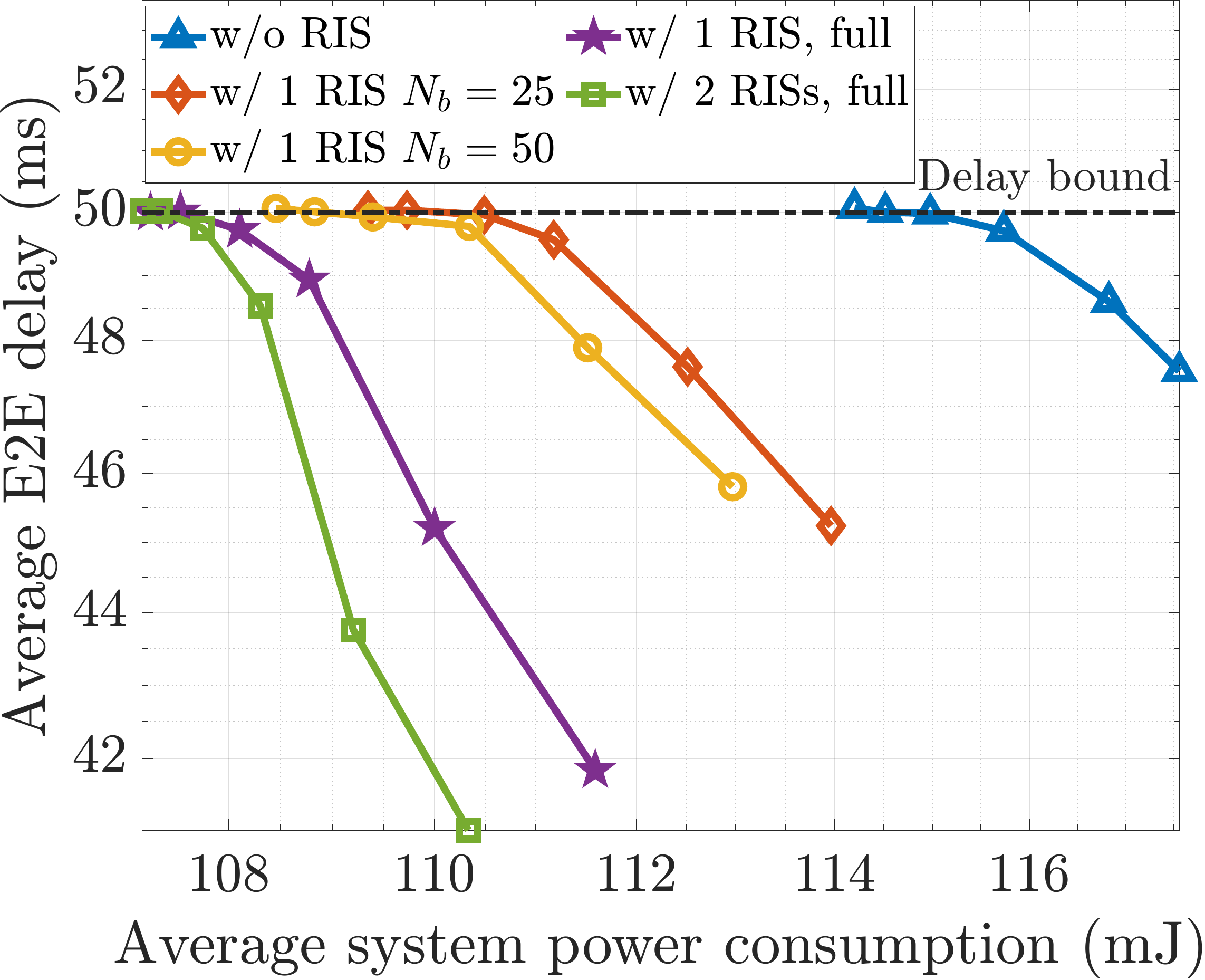}
        \label{fig:tradeoff}
    }
    \subfloat[Average users' energy vs. $V$]{
        \includegraphics[width=0.32\textwidth]{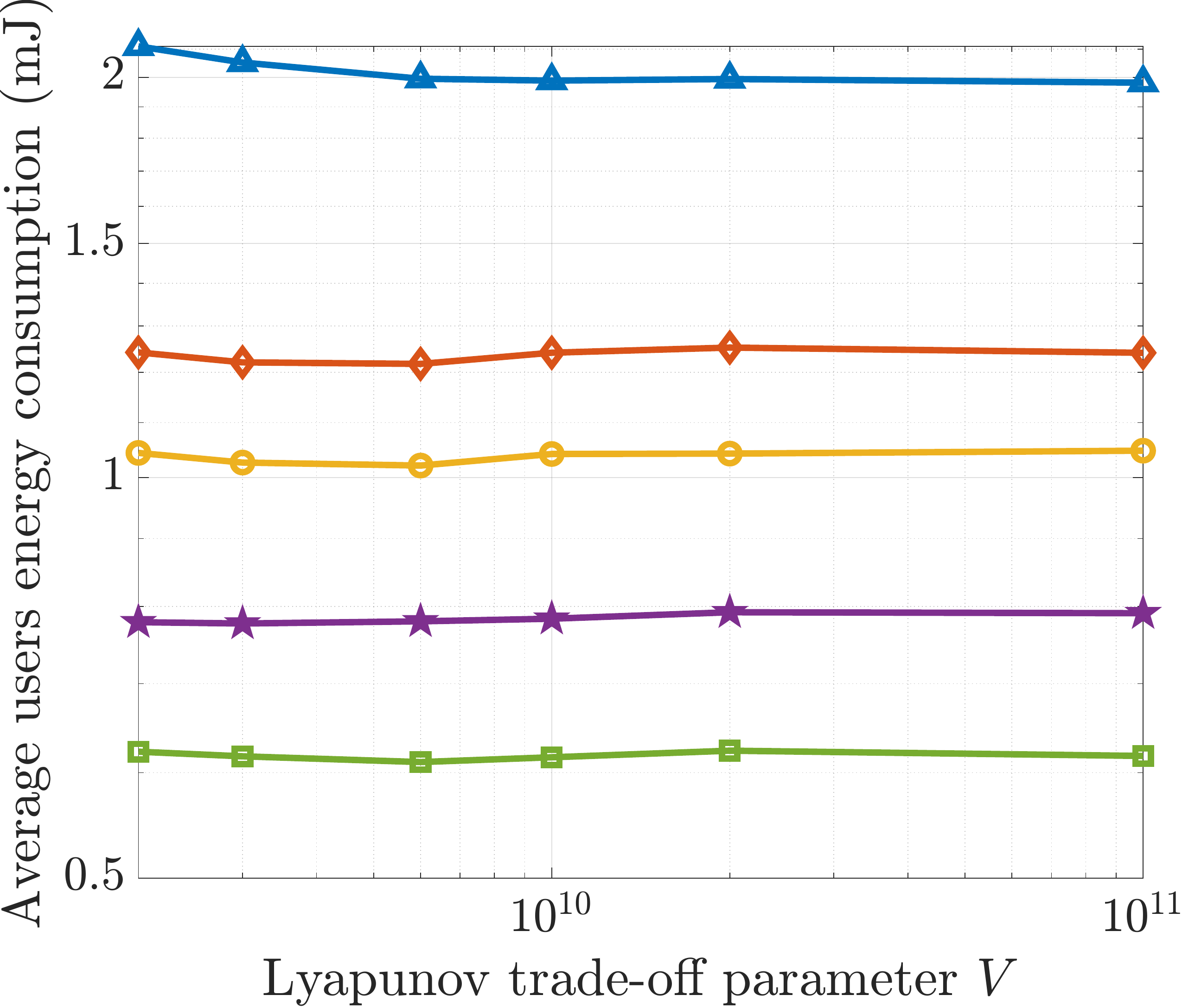}
        \label{fig:UE_energy_vs_V}
    }
    \subfloat[Average AP energy vs. $V$]{
        \includegraphics[width=0.318\textwidth]{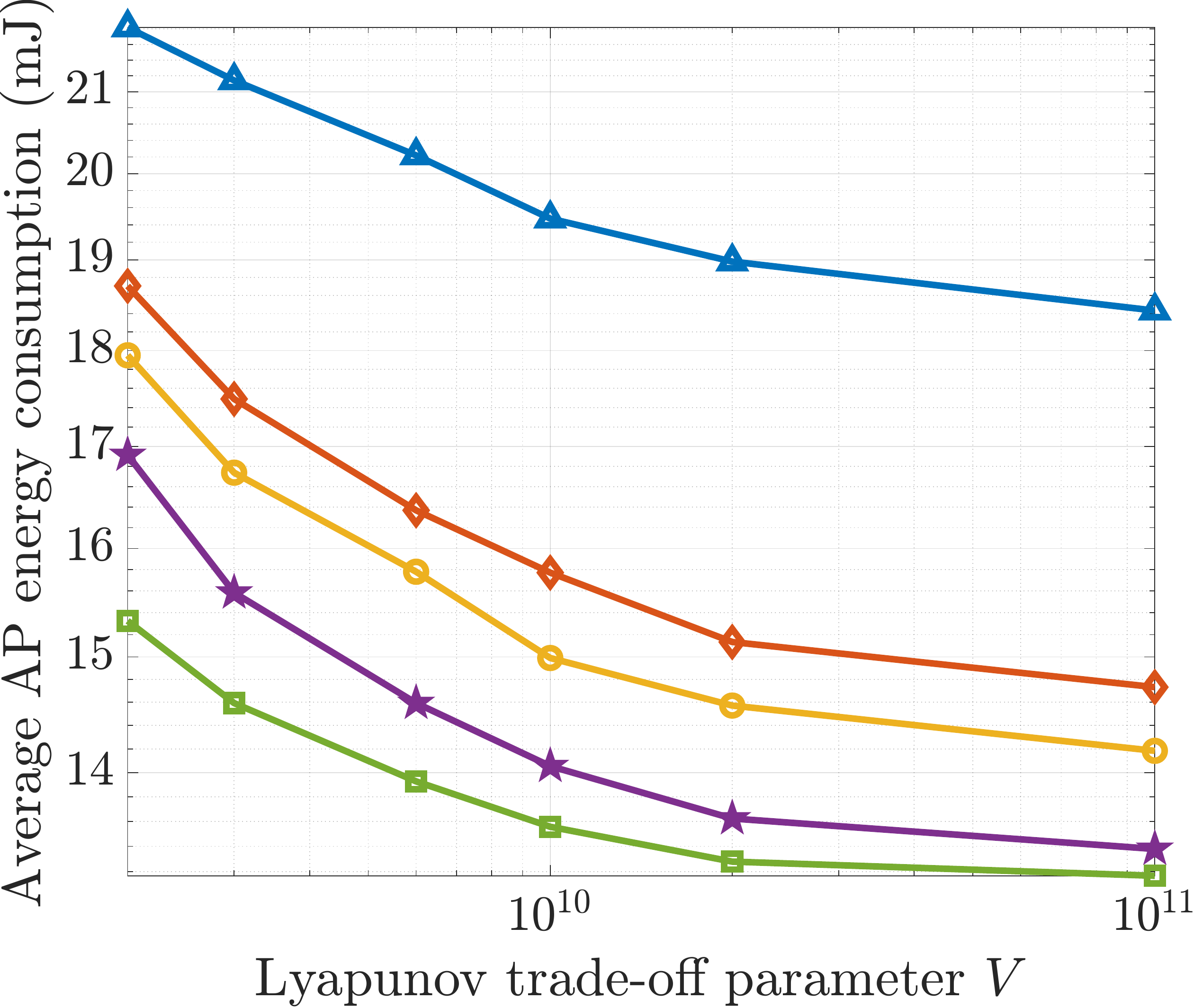}
        \label{fig:AP_energy_vs_V}
    }

    \subfloat[Average ES energy vs. $V$]{
        \includegraphics[width=0.32\textwidth]{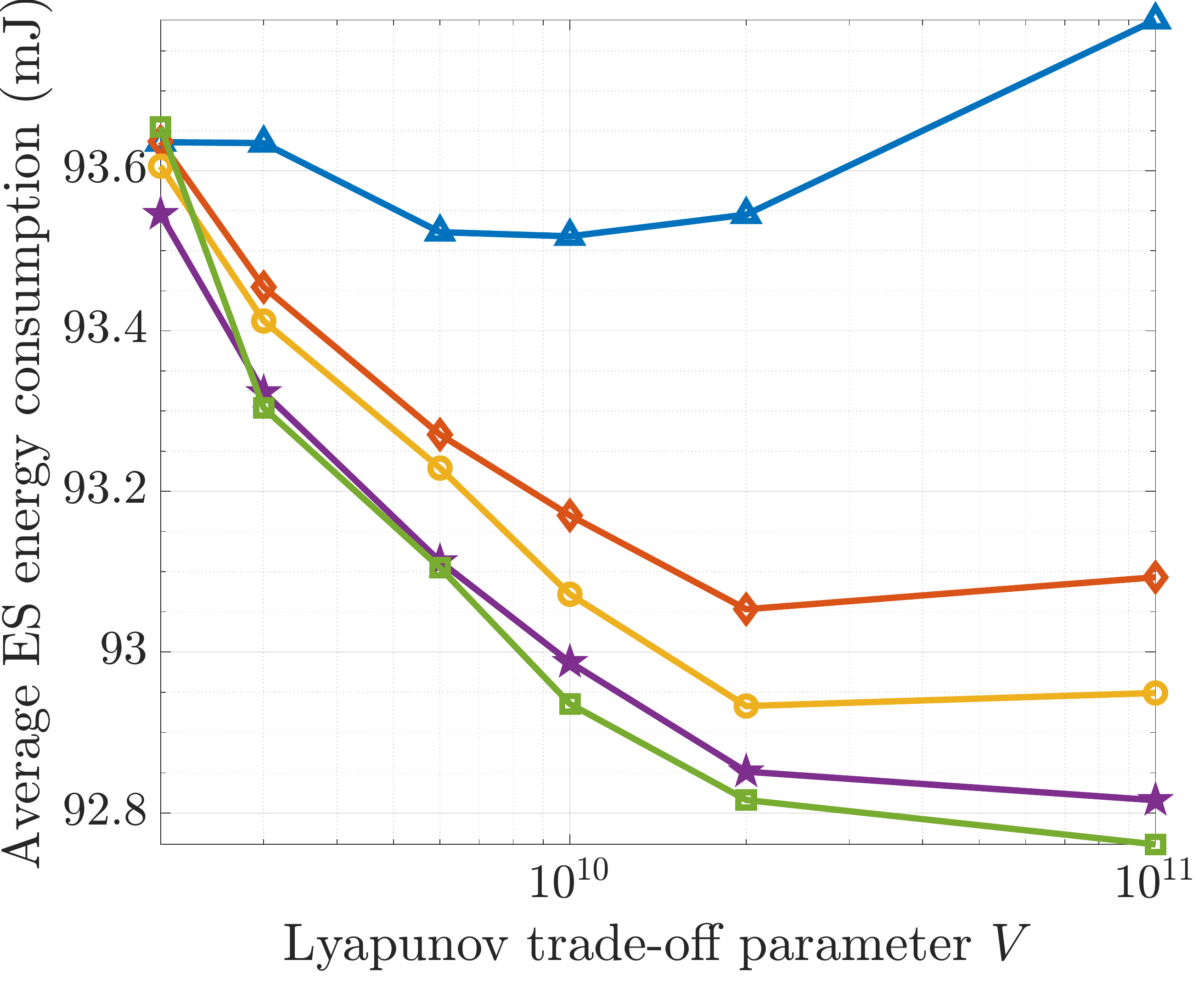}
        \label{fig:server_energy_vs_V}
    }
    \subfloat[Average RIS energy vs. $V$]{
        \includegraphics[width=0.318\textwidth]{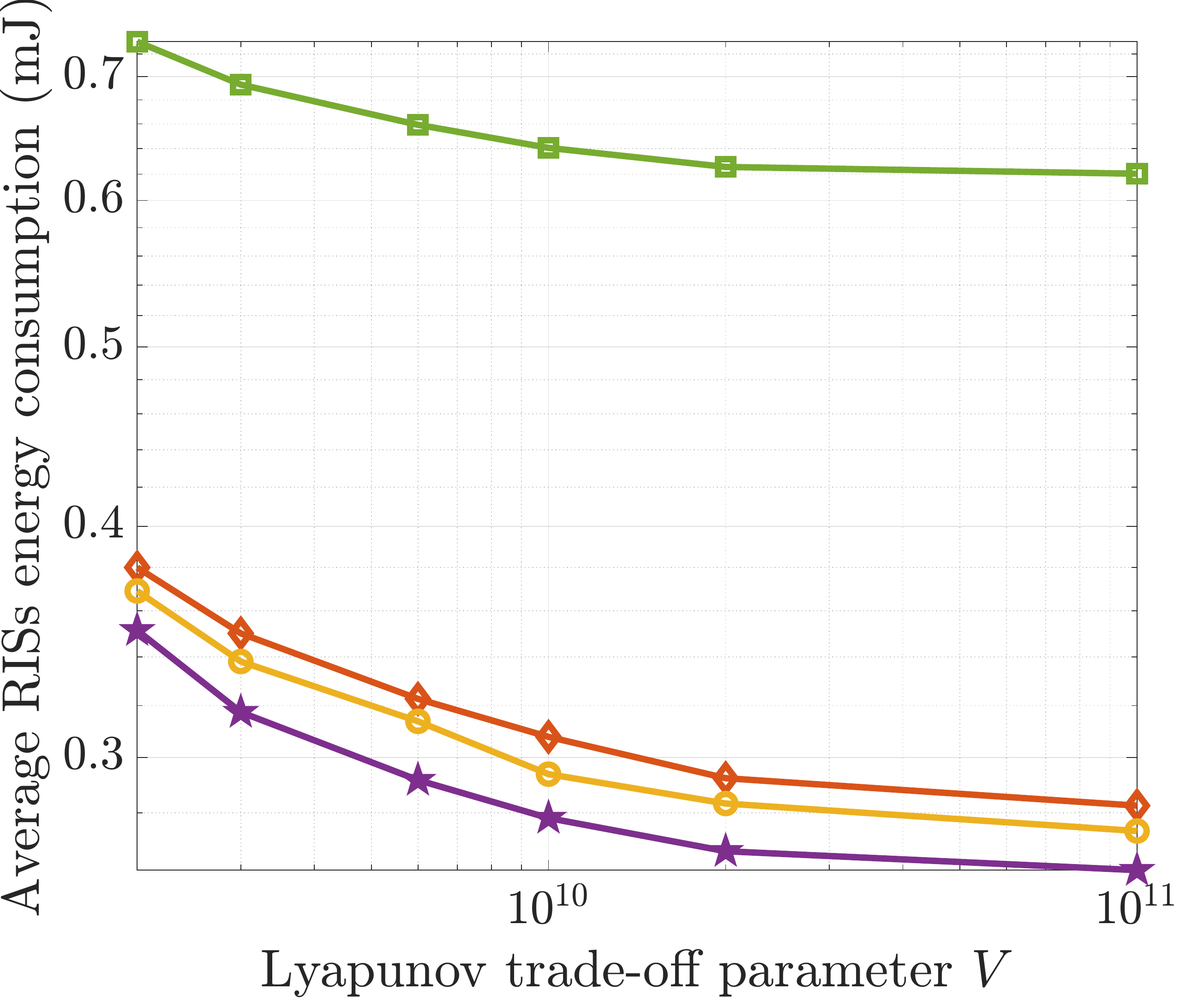}
        \label{fig:RIS_energy_vs_V}
    }
    \subfloat[AP duty cycle for different settings]{
        \includegraphics[width=0.318\textwidth]{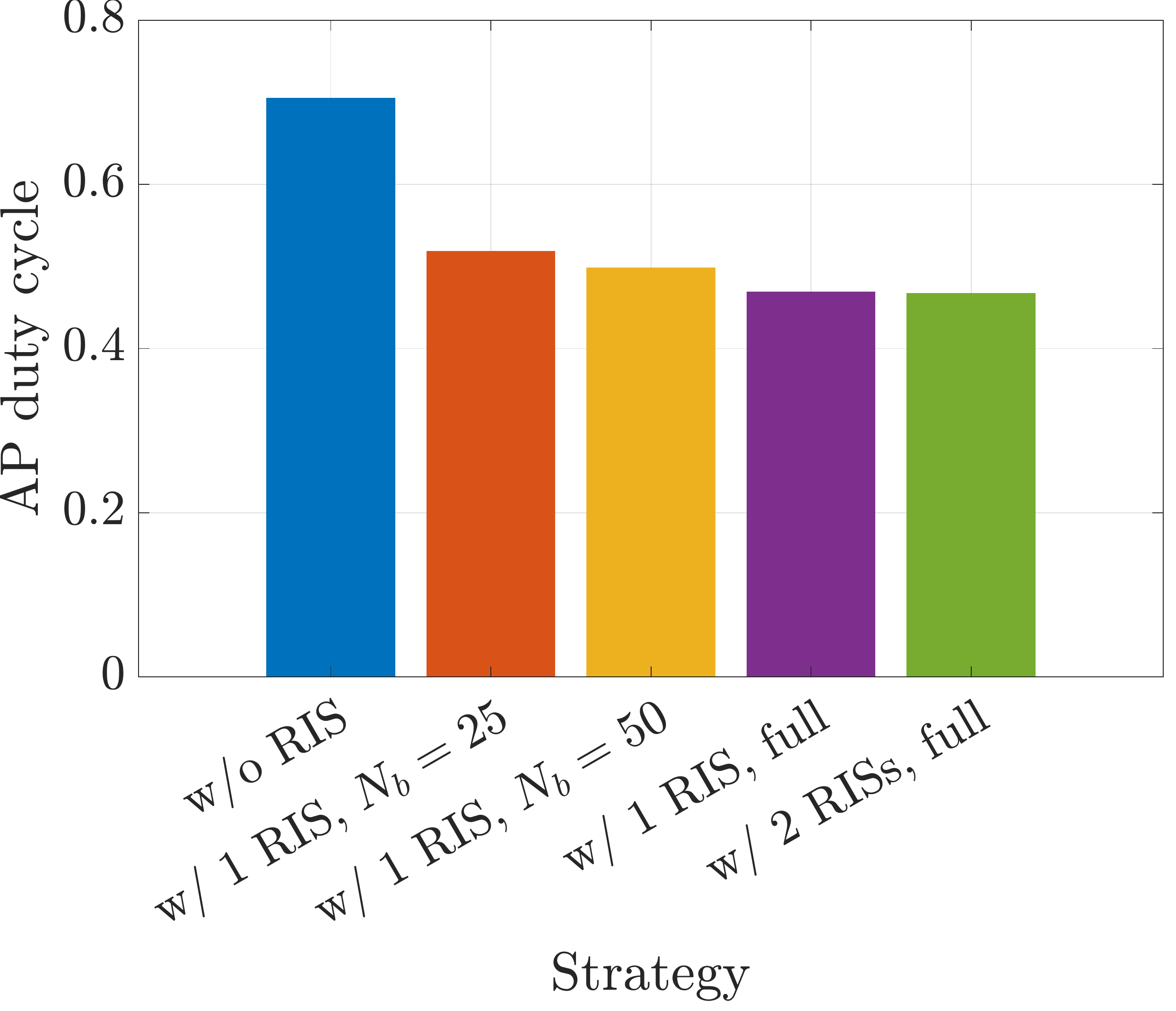}
        \label{fig:duty_cycle}
    }
    \caption{Energy-delay trade-off in a multi-user settings
    }
    \label{fig:energy_delay}
    \vspace{-0.3 cm}
\end{figure*}

\subsection{Energy-delay trade-off}

In this section, we illustrate the trade-off between user energy consumption and E2E delay, obtained with our strategy by tuning the trade-off parameter $V$ (cf. \eqref{Drift_pp}). We first present a single user setting, to then extend it to the multi-user scenario already described. For this first simulation, we consider a user-centric strategy, i.e., $\sigma=1$ (cf. \eqref{weigthed_energy}).
\subsubsection{Single user / single RIS}
We start from a simple scenario that involves the presence of one user, with an arrival rate $\bar{A}_1=500$ kbps, and (possibly) one RIS (i.e., the first ones listed in the simulation setup). In Fig. \ref{fig:tradeoff_su}, we show the E2E delay as a function of the user energy consumption, comparing a scenario without RIS with the case where one RIS is exploited, considering also imperfect channel state information (CSI). In particular, we consider both the perfect CSI case, and two cases in which the latter is estimated with an error, with $\eta$ denoting the estimation signal to noise ratio. The curves in Fig. \ref{fig:tradeoff_su} are obtained by increasing the Lyapunov trade-off parameter $V$ from right to left. As expected, by increasing $V$, the energy consumption decreases, while the average E2E delay increases up to the desired maximum bound $D_k^{\textrm{avg}}=50$ ms, for all the proposed settings. Since this work represents the first contribution on RIS-aided dynamic edge computing, the w/o RIS cases in Fig. \ref{fig:tradeoff_su} represent the current state of the art. Then, from Fig. \ref{fig:tradeoff_su}, we can notice how the proposed method exploiting RISs largely outperform the case without RIS in terms of energy-delay tradeoff. Also, the imperfect knowledge of channel states has a small impact on the performance (especially in the RIS aided scenario), thanks to the C-approximation concept introduced in Section \ref{sec_resource_all}.

\subsubsection{Multiple users / multiple RISs}

Now, we simulate a more challenging case that encompasses multiple users and possibly multiple RISs, as described in the simulation setup. For this simulation, we consider a holistic strategy that equally weights users and network energy consumption, i.e., $\sigma=0.5$ (cf. \eqref{weigthed_energy}). Thus, in Fig. \ref{fig:tradeoff}, we show the E2E delay versus the network energy consumption, considering $5$ different conditions:
\begin{itemize}
    \item A scenario without RISs;
    \item A scenario with $1$ RIS, i.e., the second RIS is switched off. Also, the optimization in Algorithm \ref{alg:RIS} is performed for each element. We term this strategy as \textit{$1$ RIS, full};
    \item A scenario with $1$ RIS, where $N_b=50$ blocks are defined, i.e., RIS element are optimized in groups of $2$. This strategy aims at reducing the complexity of the greedy strategy in Algorithm \ref{alg:RIS}.  Thus, given the number of elements $N_i$, elements are optimized, thorough Algorithm \ref{alg:RIS}, in groups of $\frac{N_i}{N_b}$ elements;
    \item A scenario with $1$ RIS, with $25$ optimization blocks, i.e. RIS elements are optimized in groups of $4$;
    \item A scenario with $2$ RISs, with full optimization.
\end{itemize}
The curves in Fig. \ref{fig:tradeoff} are obtained by increasing the Lyapunov trade-off parameter $V$ from right to left. By increasing $V$, each curve reaches a different value of the energy consumption, while converging to the desired delay bound.  As expected, all scenarios with RISs outperform the scenario without RIS, with the full optimization (with both $1$ and $2$ RISs) achieving the largest gain. The block optimization (with $N_b=25$ and $N_b=50$) reduces complexity at the cost of increased energy with respect to the full strategies.

\begin{figure}[t]
    \centering
    \includegraphics[width=7cm]{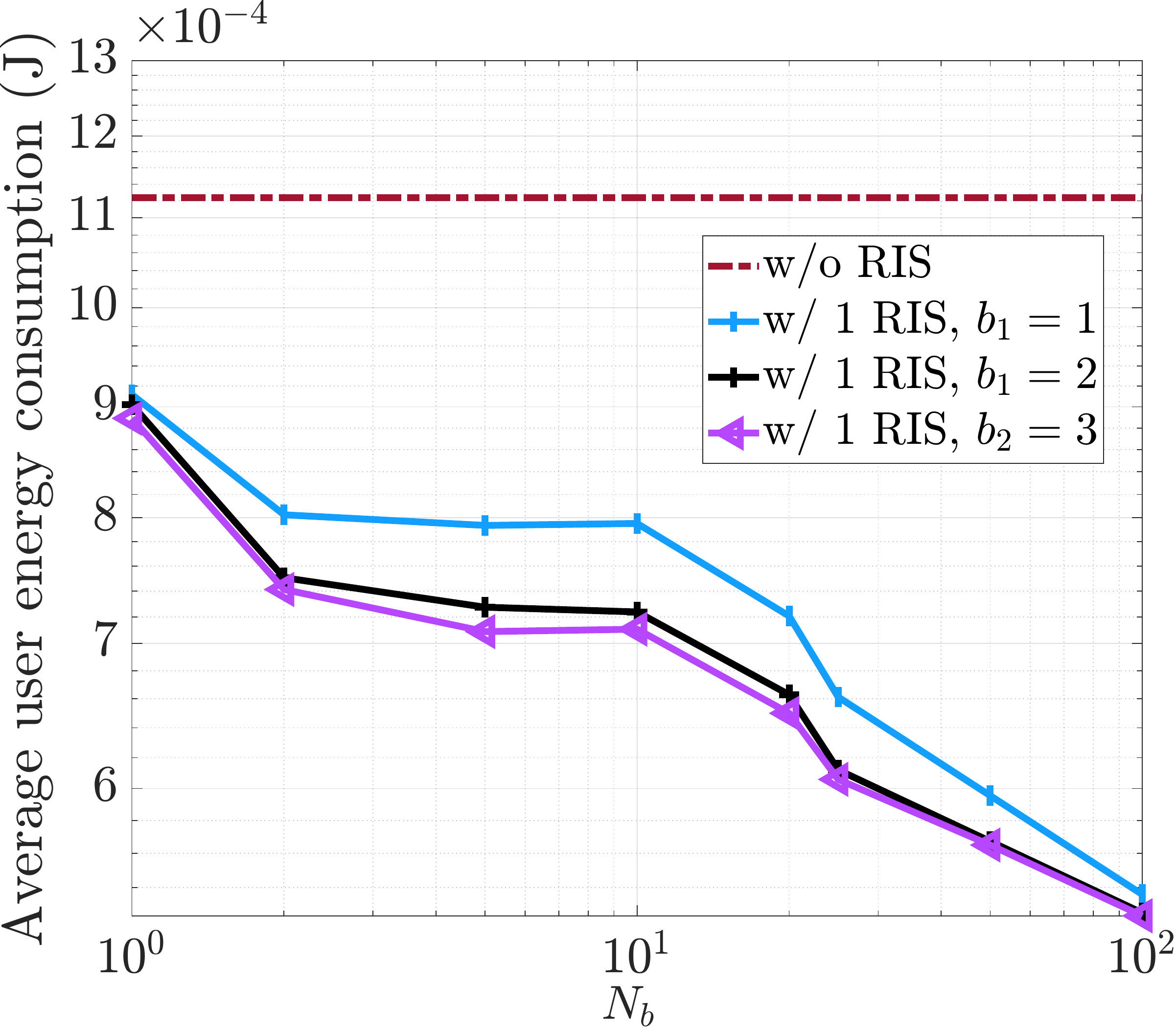}
    \caption{Average user energy consumption vs. $N_b$}
    \label{fig:user_energy_vs_blocks}
\end{figure}

By looking at Fig. \ref{fig:tradeoff}, one may conclude that the gain in terms of overall network energy consumption could be negligible. However, this is not true, since we need to analyze the single sources of energy consumption (i.e., users and network elements) separately. In particular, in Fig. \ref{fig:UE_energy_vs_V}, we show the average sum energy consumption of all users as a function of the trade-off parameter $V$, for the same values used to obtain Fig. \ref{fig:tradeoff}. Let us first notice that, while the whole network energy consumption is a monotone non-increasing function of $V$, this does not necessarily hold for the single source of energy (users, AP, ES and RISs), due to the fact that we minimize a weighted sum of the energy sources (cf. \eqref{weigthed_energy}). But most importantly, from Fig. \ref{fig:UE_energy_vs_V}, we can notice the considerable energy gain in terms of users energy consumption obtained in all the scenarios with RISs, for all values of $V$, with respect to the scenario without RIS. Also, if we concentrate on the largest value of $V$, we can compare the strategies for the  same maximum average E2E delay (i.e., the bound in Fig. \ref{fig:tradeoff}). As a result, from Fig. \ref{fig:UE_energy_vs_V}, the strategy with $2$ RISs yields a user energy consumption more than $3$ times lower than the value achievable in the non-RIS scenario. In the case of $1$ RIS optimized with $N_b=25$ elements, we obtain around a $30$\% gain. This reduced gain is the price paid by the complexity reduction with respect to the full optimization. Similar consideration can be made for the AP energy consumption in Fig. \ref{fig:AP_energy_vs_V}, which shows considerable energy gains. This is due to the fact that, since uplink and downlink communications are empowered by the RISs, the users and the AP are able to transmit more data when the AP is active, leaving more time to join the sleep state and save energy (cf. \eqref{AP_energy}). Thus, the AP duty cycle is reduced by the presence of RISs, as we can see from Fig. \ref{fig:duty_cycle}, which shows the results of the different strategies for the last value of $V=10^{11}$.
\begin{figure*}[t]
    \centering
    \subfloat[Average users energy vs. $t$]{
        \includegraphics[width=0.30\textwidth]{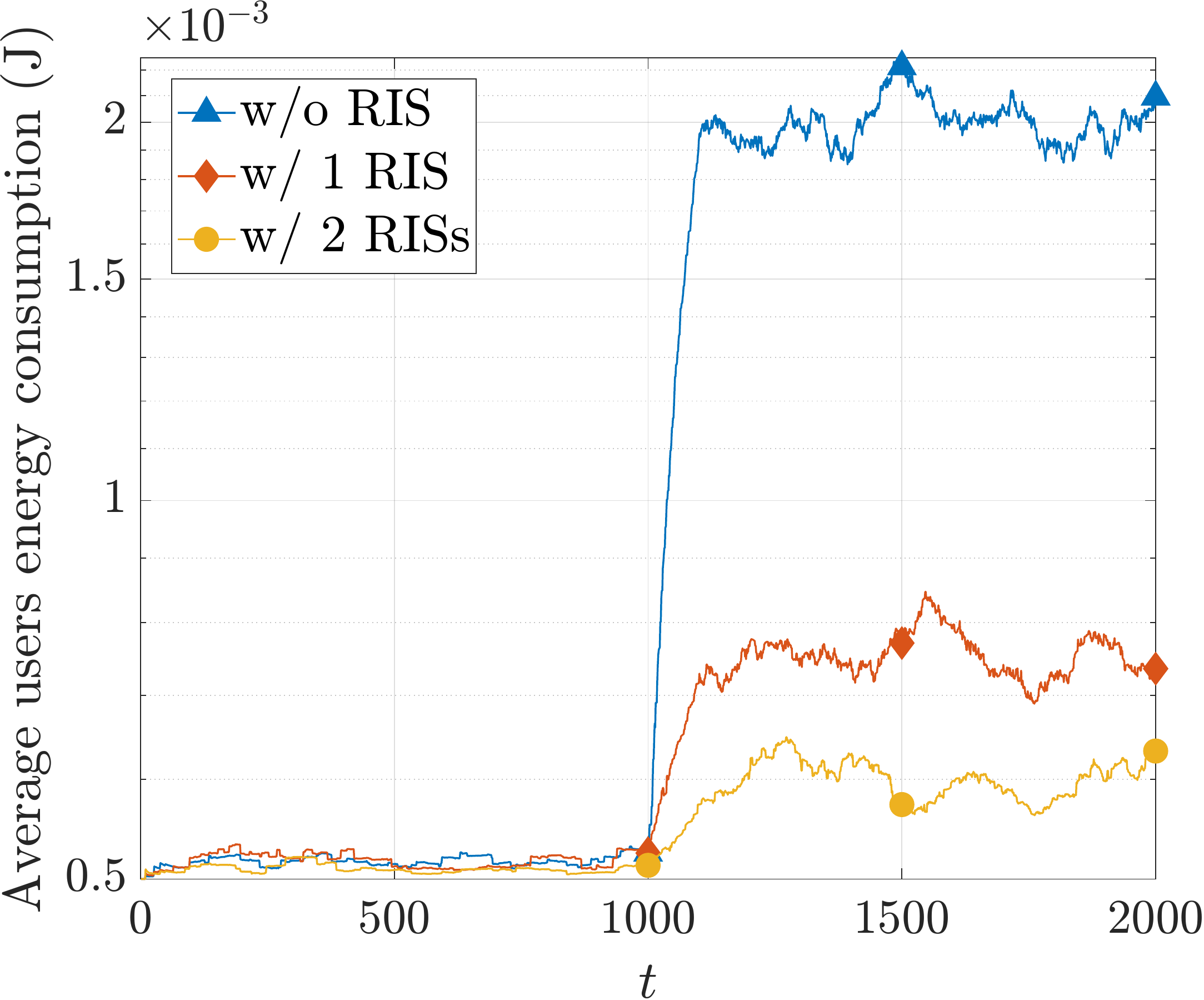}
        \label{fig:user_energy_vs_t}
    }
    \subfloat[Average system energy vs. $t$]{
        \includegraphics[width=0.30\textwidth]{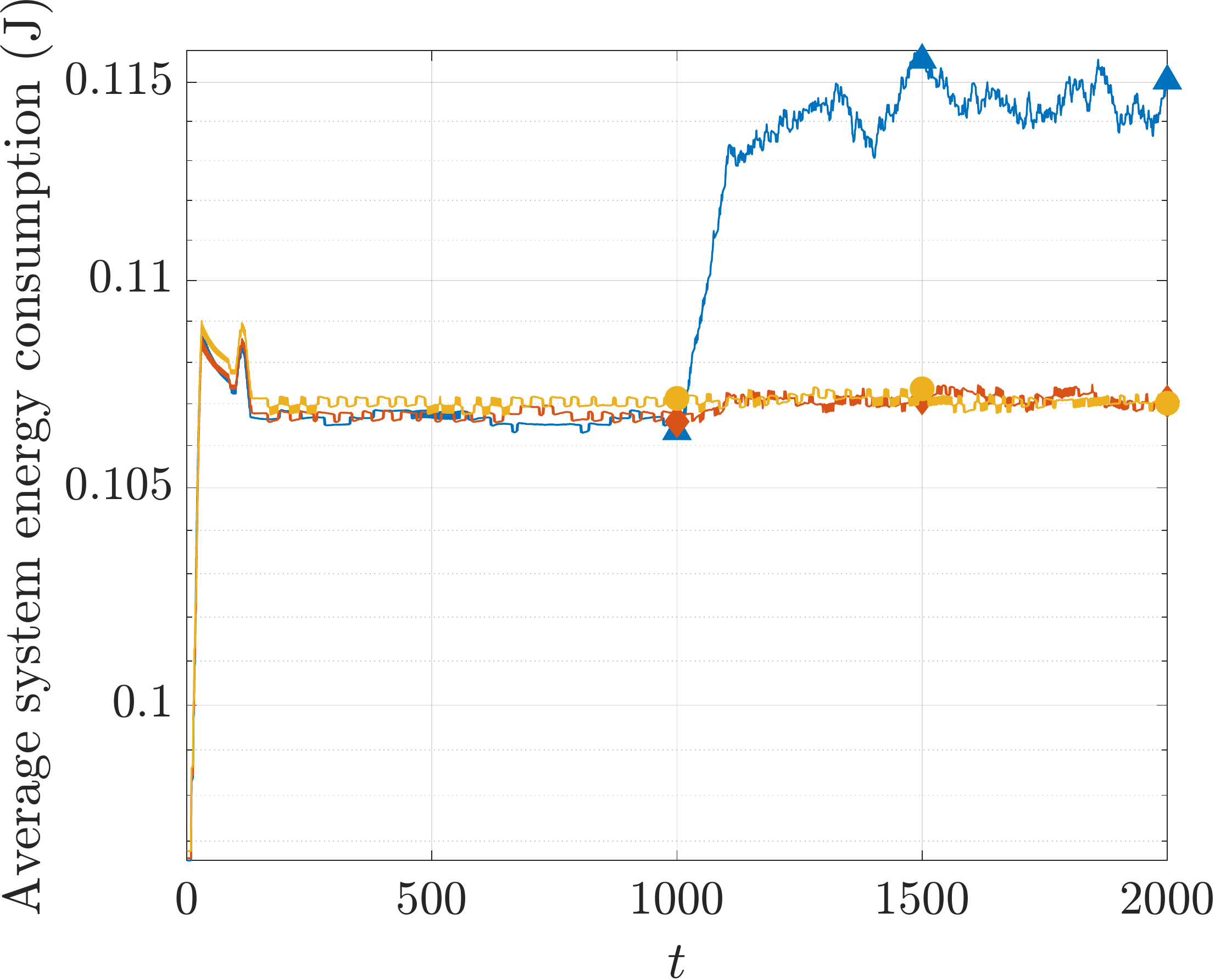}
        \label{fig:system_energy_vs_t}
    }
    \subfloat[Average delay vs. $t$]{
        \includegraphics[width=0.30\textwidth]{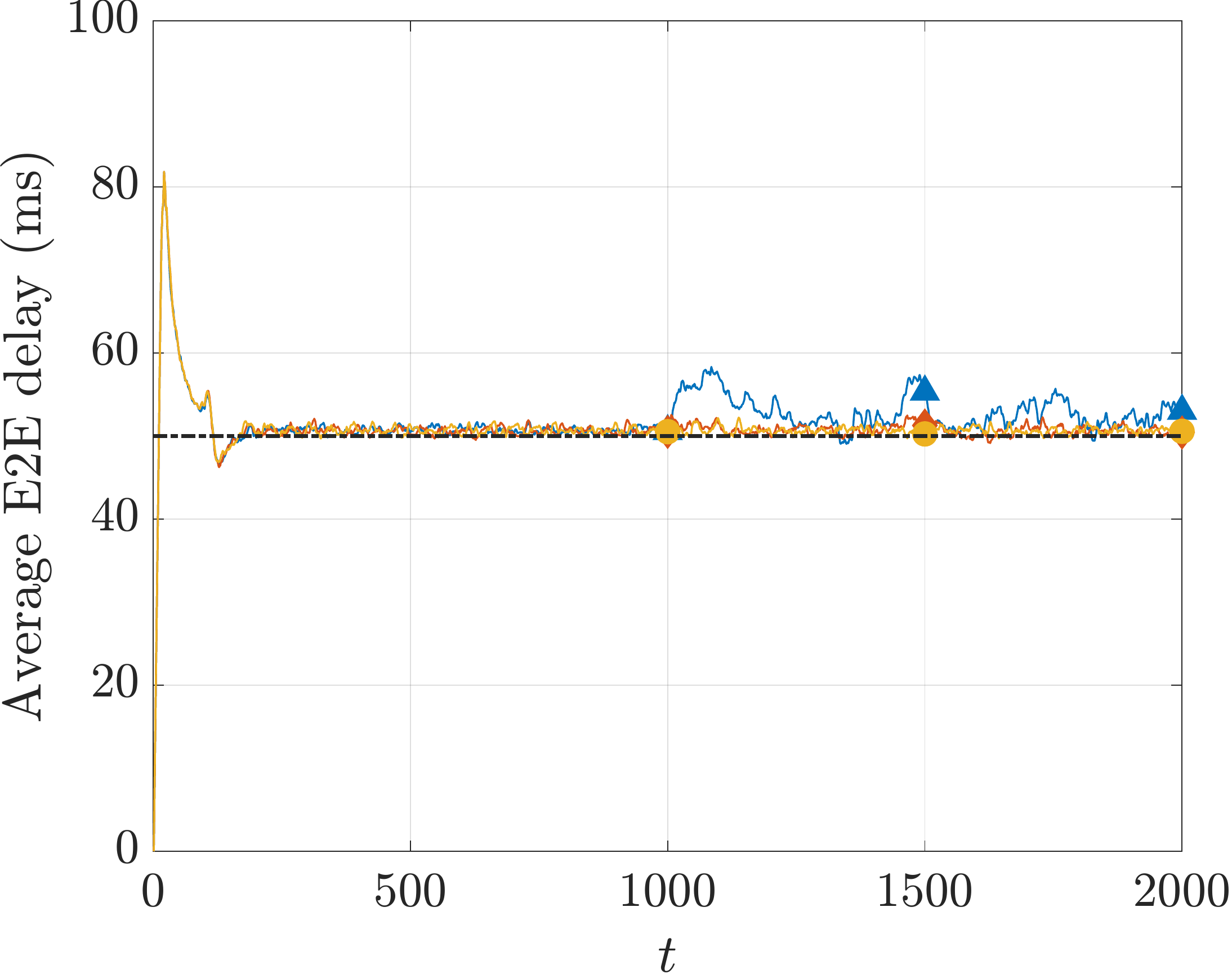}
        \label{fig:delay_vs_t}}

    \caption{Non-stationary scenario
    }
    \label{fig:non_stat_scenario}
    \vspace{-0.4 cm}
\end{figure*}
The effect of RISs is instead less visible on the energy consumption of the server, illustrated in Fig. \ref{fig:server_energy_vs_V}, which is stable around similar values for all scenarios and for all values of $V$.
\begin{table}[t]
\centering
\caption{Time efficiency gain}
\label{tab:complexity_gain}
\resizebox{.95\columnwidth}{!}{%
\begin{tabular}{ccccccccc}
\hline
\multicolumn{1}{l}{$b$/$N_b$} & 1      & 2    & 5    & 10    & 20    & 25    & 50    & 100    \\ \hline
1                             & 0.5\% & 1\% & 2\% & 4\%  & 7\%  & 8\%  & 15\% & 29\%  \\ \hline
2                             & 2\%   & 3\% & 4\% & 6\%  & 10\% & 14\% & 26\% & 50\%  \\ \hline
3                             & 1\%   & 2\% & 5\% & 10\% & 22\% & 23\% & 54\% & 100\% \\ \hline
\end{tabular}%
}
\vspace{-.4 cm}
\end{table}
Finally, the energy consumption of the RISs is shown in Fig. \ref{fig:RIS_energy_vs_V}, where we can notice the increased energy consumption with $2$ RISs. Obviously, in the case without RISs, the energy consumption is equal to zero. In summary, the take-home message of Fig. \ref{fig:energy_delay} is three-fold:
\begin{itemize}
    \item Our dynamic strategy is able to reduce the whole energy consumption, with the cost of an increased delay, up to the threshold defined through constraint $(a)$ of \eqref{Problem};
    \item Empowering MEC with RISs slightly reduces the whole network energy (a non-straightforward fact due to the presence of the RIS energy consumption), while it yields a large gain in terms of users and AP energy consumption.
    \item The complexity of Algorithm \ref{alg:RIS} can be reduced by optimizing groups of elements, with the cost of a decreased (yet considerable) gain in terms of energy performance.
\end{itemize}

\vspace{-.35 cm}
\subsection{User-centric optimization with different blocks $N_b$} The results obtained in Fig. \ref{fig:energy_delay} motivate us to explore the performance in terms of energy consumption and complexity in the user-centric case (i.e., $\sigma=1$ in \eqref{weigthed_energy}), by varying the number of blocks $N_b$ and the number of bits $b_i$ used to optimize RIS's phases (cf. \eqref{phases}). To this aim, in Fig. \ref{fig:user_energy_vs_blocks}, we illustrate the users energy consumption as a function of the number of blocks $N_b=[1,2,5,10,20,25,50,100]$. Let us recall that $N_b=100$ corresponds to the full optimization of Fig. \ref{fig:tradeoff}, while $N_b=1$ is the lowest complexity strategy, since it excites all RIS elements with the same phase. For this simulation, we consider only one RIS, and we compare the results with the non-RIS scenario, which is depicted with a horizontal line. From Fig. \ref{fig:user_energy_vs_blocks}, we can notice how, using the RIS is always beneficial, even in the case with $N_b=1$, although with a slight gain with respect to the non RIS scenario. As expected, by increasing the number $N_b$ of blocks, the energy consumption decreases thanks to the larger degrees of freedom in optimizing the RIS elements. Also, increasing the number of bits yields a further reduction in the energy consumption, which is more or less appreciable depending on $N_b$. Finally, from a complexity point of view, we show in Table \ref{tab:complexity_gain} the percentage of saved time in running a single instance of Algorithm \ref{alg:RIS}, with respect to the highest complexity strategy ($b_1=3$, $N_b=100$). From Table \ref{tab:complexity_gain}, decreasing $b_i$ as well as $N_b$, we can achieve a considerable gain in terms of computation time needed to find the solution, paid with an increased energy consumption. This quantifies the inherent energy-complexity trade-off introduced by the RIS block optimization.
\vspace{-.35 cm}

\subsection{Adaptation in non-stationary scenarios}

As a final result, we illustrate how the proposed method behaves in a non-stationary scenario with dynamic channel blocking. We assume that, at the beginning of the optimization, no obstacle obscures the direct path between AP and users. Then, at slot number $1000$, an obstacle with $30$ dB attenuation is interposed in the direct path. For this simulation, we consider again a holistic strategy ($\sigma=0.5$), and we compare the results without RIS, with $1$ RIS and with $2$ RISs, with full RIS optimization ($N_b=100$). Then,   in Fig. \ref{fig:user_energy_vs_t}, \ref{fig:system_energy_vs_t} and \ref{fig:delay_vs_t}, we illustrate the moving average of the users energy consumption, system energy consumption, and average E2E delay, obtained by averaging these quantities over the last $100$ slots. From Figs. \ref{fig:user_energy_vs_t} and \ref{fig:system_energy_vs_t}, we can notice how, at the beginning, all scenarios converge to a similar user and system energy consumption, due to the fact that the direct path is in good conditions and the RIS does not yield considerable gains. However, when the blockage occurs, the case without RIS is heavily affected from a user energy consumption perspective. This is due to the fact that the direct path is strongly attenuated, which requires higher user transmit power and more AP activity to cope with the arrival rate and stabilize the system. On the contrary, the presence of a RIS determines only a mild effect of a blocking event on the performance. Indeed, as we can see from figs. \ref{fig:user_energy_vs_t} and \ref{fig:system_energy_vs_t}, with one RIS, the energy consumption is affected due to the blocked direct path, but it is able to converge (in a few time-slots) to a new value much lower than the non-RIS case thanks to the alternative path and the inherent gain of the RIS channel. With two RISs, this gain is even more visible.  Finally, from Fig. \ref{fig:delay_vs_t}, we can notice how the delay stabilizes, in both cases, around the threshold, albeit a slight violation caused by the fact that the average is performed over a small window of $100$ slots.


\vspace{-.2cm}
\section{Conclusions}

In this paper, we have proposed a novel algorithm for energy-efficient, low-latency dynamic edge computing, empowered with reconfigurable intelligent surfaces. The method hinges on stochastic optimization tools, learning dynamically and jointly the phases of RISs elements, the radio parameter of users and of the access point (i.e., powers and active states), and the CPU frequencies of the edge server. Even in the complex dynamic MEC scenario considered in the paper, the proposed approach requires only low-complexity procedures at each time slot, and enables online adaptation of the RISs configuration to dynamically shape the wireless propagation channel. Being fully adaptive, the method does not need any apriori knowledge of channel and data arrival statistics. Numerical results assess the performance of the proposed strategy, illustrating the potential gain and adaptation capabilities achievable endowing MEC systems with multiple reconfigurable intelligent surfaces.
\vspace{-.35 cm}
\bibliographystyle{IEEEbib}
\bibliography{refs}

\end{document}